\documentclass[twocolumn]{article}
\usepackage{graphicx} % Required for inserting images
\usepackage{hyperref}
\usepackage{amsthm}
\usepackage{authblk}
\usepackage{multicol}
\usepackage{amssymb}
\usepackage{amsmath}
\usepackage{lipsum}
\usepackage{cuted} 
\usepackage{afterpage}
\usepackage{paracol}
\usepackage{cutwin}
\usepackage{mathrsfs}
\usepackage[numbers,square]{natbib} 
\usepackage{fancyhdr}
\usepackage{subfig}
\newcommand{\mywide}[2][0.9]{\begin{strip}
\noindent\parbox{\textwidth}{#2}
\end{strip}}

\title{Energy Balance of a Boson Gas at Zero Temperature in Curved Spacetime}

\author[1]{Jorge Meza-Domínguez\thanks{E-mail: \href{mailto:jorge.meza@cinvestav.mx}{jorge.meza@cinvestav.mx}}}
\author[2]{Tonatiuh Matos\thanks{E-mail: \href{mailto:tonatiuh.matos@cinvestav.mx}{tonatiuh.matos@cinvestav.mx}}} 
\author[3]{Pierre-Henri Chavanis\thanks{E-mail: \href{mailto:chavanis@irsamc.ups-tlse.fr}{chavanis@irsamc.ups-tlse.fr}}}
\affil[1,2]{Departamento de F\'{\i}sica, Centro de Investigaci\'on y de Estudios Avanzados del Intituto Politécnico Nacional, Av. Intituto Politécnico Nacional 2508, San Pedro Zacatenco, M\'exico 07360, CDMX.}
\affil[3]{Laboratoire de Physique Th\'eorique, Universit\'e Paul Sabatier, 118 Route de Narbonne, 31062 Toulouse Cedex 9, France}
\date{}
\begin{document}
\maketitle
\mywide{
\begin{abstract}
We develop a comprehensive thermodynamic description for a zero-temperature boson gas in a fixed, classical curved spacetime, integrating energy conservation with information-theoretic principles. Using the hydrodynamic Madelung representation within the ADM formalism, we establish two fundamental relationships: an energy balance equation representing the first law of thermodynamics from a spacetime perspective, and an information-theoretic constraint connecting Fisher entropy to the dynamical evolution of the boson density. This formulation clearly separates energy transport from the conservation of quantum information encoded in the boson gas, while revealing how such information is preserved in curved backgrounds. The introduction of a stochastic velocity provides a bridge between quantum potential effects and underlying spacetime fluctuations. We demonstrate the consistency of our framework through detailed analyses of quantum systems in both Minkowski and Schwarzschild spacetimes. This work provides a unified foundation for studying relativistic bosonic systems, with direct relevance to boson stars and scalar field dark matter models.\\
\textbf{Keywords:} Boson Gas, Curved Spacetime, Thermodynamics, Madelung Transformation, Geodesic Velocity, Quantum Potential, Fisher Information, Energy Balance.
\end{abstract}
}

\section{Introduction}

The reconciliation of thermodynamics with general relativity remains a pivotal challenge in theoretical physics \cite{tHooft2009}. In curved spacetime, where the metric is a dynamical entity, concepts such as local energy density and global conservation laws become inherently ambiguous \cite{Alcubierre2008}. This ambiguity complicates the formulation of a consistent thermodynamic description for relativistic systems, from self-gravitating boson stars to scalar field dark matter models \cite{Guzmn2000, Matos2001, Chavanis2011,Chavanis2025}. A crucial step forward is the establishment of a clear energy balance equation that accounts for the diverse energy contributions within a general relativistic framework.

Scalar fields offer a versatile approach to modeling such systems. Described fundamentally by the Klein-Gordon equation and, when coupled to electromagnetism, by Maxwell's equations, they admit a hydrodynamic reformulation via the Madelung transformation \cite{Madelung1927}. This transformation recasts the wave equations into a fluid-like description, revealing an underlying quantum potential \cite{Simeonov2025}. While this leads to the Gross--Pitaevskii equation in flat spacetime, its extension to curved backgrounds is essential for astrophysical and cosmological applications.

In a recent contribution, Matos et al.~\cite{Matos2019} established the thermodynamic foundation for bosonic gases in curved spacetime. Their approach combined the ADM $3+1$ decomposition \cite{Alcubierre2008} with the Madelung transformation \cite{Madelung1927} of the Klein-Gordon-Maxwell system \cite{Greiner2000, Chavanis2017,Chavanis2017_b,Sin1992}, identifying distinct energy components (kinetic, quantum, gravitational, and electromagnetic) and unifying them into a single balance equation. This marked a significant advance toward formulating the first law of thermodynamics for scalar fields in general relativity.

However, a single unified equation, while complete, intertwines multiple physical processes. A more nuanced thermodynamic understanding often requires distinguishing between the conservation of a quantity along the fluid flow and its transport across spacetime \cite{Jimnez-Aquino2018}. For instance, relativistic hydrodynamics separately treats energy-momentum conservation and particle number current. Similarly, decomposing the energy balance into complementary equations could yield deeper insight into the system's dynamics.

In this work, we build upon the foundation of Matos et al.~\cite{Matos2019} by enriching the thermodynamic description with insights from information theory and stochastic mechanics. Using the same ADM $3+1$ framework \cite{Alcubierre2008} and hydrodynamic formulation \cite{Madelung1927, Simeonov2025}, we demonstrate that the incorporation of Fisher information concepts \cite{Frieden2004, DnesPetz2008} and stochastic dynamics \cite{Nelson1966, delaPea2015} leads to a more fundamental decomposition of the system's thermodynamic structure. This enhanced perspective reveals that the dynamics can be elegantly captured through two fundamental equations:

\begin{itemize}
    \item An energy balance equation governing the flux of total energy and its coupling to changes in the self-interaction potential, and
    \item An information-theoretic constraint linking the Fisher entropy~\cite{Freidel2008, DnesPetz2008, Caticha2011, Braunstein1994} to the spacetime evolution of the boson density and the self-interaction potential.
\end{itemize}

This dual formulation not only clarifies how energy is conserved from the perspective of a comoving observer and transferred from that of a static observer, but also reveals how quantum information---encoded in the Fisher entropy~\cite{Frieden2004, DnesPetz2008} and connected to holographic principles~\cite{Freidel2008}---is dynamically conserved in curved spacetime. The introduction of a stochastic velocity~\cite{Nelson1966, Nelson2012, delaPea1996}\footnote{We use the term ``stochastic velocity'' for $u_{\mu}$ following the terminology of stochastic mechanics \cite{Nelson1966,Nelson2012}. In the present hydrodynamic formulation, $u_{\mu} = (\hbar/2m)\nabla_{\mu}\ln n$ serves as a technical device to separate the diffusive and conservative components of the dynamics, without implying any commitment to a particular interpretation of quantum mechanics.} further bridges the quantum potential with underlying spacetime fluctuations.

The ambiguity in defining local energy and global conservation laws in general relativity suggests that is possible gravity may possess a fluctuating nature at a fundamental scale \cite{Jacobson1995}. This perspective aligns naturally with approaches in stochastic quantum gravity \cite{Hu2008,Padmanabhan2010}, where spacetime itself is subject to random fluctuations, possibly linked to a background of gravitational waves or non-deterministic gravitational degrees of freedom \cite{Escobar-Aguilar2025, Gallegos2021}. Our introduction of a stochastic velocity in the Madelung hydrodynamic formulation thus acquires deeper relevance. In particular, a stochastic background of gravitational waves would locally modulate the causal structure and curvature, inducing random variations in the Lie derivative of the energy density and in the evolution of the Fisher entropy. Thus, the energy balance equation and information-theoretic constraint we derive could be interpreted as equations averaged over stochastic realizations of the gravitational field.

We validate this approach through concrete examples---including the harmonic oscillator~\cite{Schiff1968}, the hydrogen atom~\cite{Schiff1968}, and the Klein--Gordon field in Schwarzschild geometry~\cite{Li2019, Hawking1975}---demonstrating its consistency and physical relevance. By integrating general relativity~\cite{Alcubierre2008}, thermodynamics~\cite{Matos2019}, quantum information theory~\cite{Frieden2004, DnesPetz2008}, and insights from stochastic quantum gravity, this work provides a refined foundation for modeling boson stars~\cite{Chavanis2011}, dark matter~\cite{Matos2001, Guzmn2000}, and other relativistic bosonic systems where quantum, gravitational, and informational aspects are inextricably intertwined.

\section{Klein-Gordon-Maxwell Equations}

The physical system under consideration consists of a charged boson gas, represented by a complex scalar field with self-interactions. The field couples minimally to electromagnetism through a gauge vector field, while gravitational interactions are incorporated via a curved spacetime geometry, consistent with the tenets of General Relativity \cite{Alcubierre2008}.

Our approach operates within a fixed, arbitrary curved spacetime background, thus circumventing the explicit use of the Einstein field equations while maintaining full general covariance. The core of our formulation lies in extending the Klein-Gordon-Maxwell system---governed by a local \(U(1)\) symmetry and well-established in flat spacetime \cite{Greiner2000}---to this general geometric setting. The spacetime structure is defined by a metric \(g\) on a 4-dimensional manifold, following the ADM decomposition formalism \cite{Alcubierre2008}.

The dynamics is governed by a gauged generalization of the d'Alembert operator, defined as \(\Box_{\rm E}=(\nabla^{\mu}+i\frac{e}{\hbar}A^{\mu})(\nabla_{\mu}+i\frac{e}{\hbar}A_{\mu})\),\footnote{The operator $\Box_E$ generalizes the d'Alembertian for charged fields in curved spacetime, preserving $U(1)$ gauge covariance.} where \(e\) denotes the coupling constant (charge) and \(A_{\mu}\) the electromagnetic 4-potential. This covariant formulation extends previous work on bosonic systems in curved spacetime \cite{Chavanis2017_b, Matos2019}. The full set of field equations is given by the Klein-Gordon equation
\begin{equation}
    \label{eqn:1}
    \Box_E \Phi - \dfrac{d V}{d \Phi^*} =0,
\end{equation}
for the complex scalar field $\Phi(x,t)$, $\Phi^*(x,t)$ is the complex conjugate, coupled to the Maxwell equations
\begin{equation}
\label{eqn:2}
    \nabla_{\mu}F^{\mu \nu}=J^{E\mu},
\end{equation}
where $J^{E\mu} = i\frac{e}{\hbar} [\Phi^* (\nabla^\mu + i\frac{e}{\hbar} A^\mu)\Phi - \Phi(\nabla^\mu - i\frac{e}{\hbar} A^\mu)\Phi^*]$ is the electric current density associated with the charged scalar field \cite{Greiner2000}. The Faraday tensor is given by
\begin{equation}
    \label{eqn:3}
    F_{\mu \nu}=\nabla_\mu A_{\nu} -\nabla_{\nu} A_{\mu},
\end{equation}
We introduce the self-interaction through the potential
\begin{equation}
    \label{eqn:4}
    V=2\frac{m^2}{\hbar^2}|\Phi|^2 \mathcal{A},
\end{equation}
describing the system, where $\mathcal{A}(x,t)$ is the effective self-interaction potential of the bosonic system, expressed in a single macroscopic ground state. This form of self-interaction potential follows established treatments of bosonic gases in curved spacetime~\cite{Chavanis2017, Matos2019}.

We employ the $3+1$ foliation metric, with the line element given by
\begin{equation}
    \label{eqn:5}
    ds^2=-Nc^2dt^2+\gamma_{ij}\left(dx^i+N^icdt\right)\left(dx^j+N^jcdt\right),
\end{equation}
where \(N(t,\mathbf{x})\) denotes the lapse function, governing the proper time interval between adjacent spatial slices for Eulerian observers. The shift vector \(N^{i}(t,\mathbf{x})\) describes the relabeling of spatial coordinates between successive hypersurfaces, while \(\gamma_{ij}(t,\mathbf{x})\) represents the induced spatial metric on each three-dimensional hypersurface, encoding the intrinsic geometry. This ADM decomposition~\cite{Alcubierre2008} not only facilitates a canonical treatment of the gravitational field but also provides a natural framework for tracking the dynamical evolution of fields and matter sources within the spatial manifold, following the standard approach in numerical relativity~\cite{Alcubierre2008}.
\section{Hydrodynamic form equations}
We define the polar representation of $\Phi$ by
\begin{equation}
    \label{eqn:6}
    \Phi = \sqrt{n}e^{i\theta},
\end{equation}
known as the Madelung transformation~\cite{Madelung1927}. Here $n$ is interpreted as the density of the boson gas and $\theta$ is the phase of the state, following the hydrodynamic formulation of quantum fields in curved spacetime~\cite{Chavanis2017, Matos2019}.

We computed the following results:
\begin{equation}
    \label{eqn:7}
    \nabla_{\mu}\Phi=\left(\frac{\nabla_{\mu}n}{2\sqrt{n}}+i\sqrt{n}\nabla_{\mu}\theta\right)e^{i\theta},
\end{equation}
\begin{equation}
    \label{eqn:8}
    \begin{split}
    \nabla^{\mu}\nabla_{\mu}\Phi= & \Big[\frac{\nabla_{\mu}\nabla^{\mu}n}{\sqrt{n}}-\frac{\nabla^{\mu}n\nabla_{\mu}n}{n\sqrt{n}}+i\frac{\nabla_{\mu}n}{\sqrt{n}}\nabla^{\nu}\theta \\
    &+i\sqrt{n}\nabla^{\mu}\nabla_{\mu}\theta-\sqrt{n}\nabla^{\mu} \theta \nabla_{\mu} \theta \Big]e^{i\theta},
    \end{split}
\end{equation}

Substituting into (\ref{eqn:1}), we obtain two equations after separating real and imaginary parts:
\begin{equation}
    \label{eqn:9}
    \frac{\Box \sqrt{n}}{\sqrt{n}} -\left(\nabla^{\mu}\theta+2\frac{e}{\hbar}A^{\mu}\right)\nabla_{\mu}\theta-\frac{e^2}{\hbar^2}A^2-\frac{2m^2}{\hbar^2}\mathcal{A}=0,
\end{equation}
\begin{equation}
    \label{eqn:10}
    \frac{\nabla_{\mu}n\nabla^{\mu}\theta}{n}+\Box \theta + \frac{e}{\hbar} \nabla_{\mu} A^{\mu}+\frac{e}{\hbar}A^{\mu}\frac{\nabla_{\mu}n}{n}=0,
\end{equation}

Equation (\ref{eqn:9}) governs the real part of the evolution of the state $\Phi$, representing the quantum Hamilton-Jacobi equation, while (\ref{eqn:10}) governs the imaginary part, corresponding to the continuity equation in the hydrodynamic formulation~\cite{Chavanis2017, Matos2019}. Note that both density and phase appear in both equations, reflecting the coupled nature of the hydrodynamic description in curved spacetime.
\\
We introduce the 4-vector
\begin{equation}
    \label{eqn:11}
    \pi_{\mu}:=\frac{\hbar}{m}\left(\nabla_{\mu}\theta+\frac{e}{\hbar}A_{\mu}\right).
\end{equation}
defined as the geodesic velocity\footnote{The term ``geodesic velocity'' refers to the mathematical structure of Eq. (\ref{eqn:17}): $\pi_\mu$ satisfies a forced geodesic equation. In the Madelung hydrodynamic formulation, $\pi_\mu$ represents the velocity field of the probability flow, not individual particle trajectories. In the absence of quantum potential, self-interactions, and electromagnetic fields, this reduces to the geodesic equation for the probability current.}\footnote{The geodesic velocity $\pi_\mu$ incorporates quantum effects through the phase $\theta$ and electromagnetic effects via $A_\mu$, generalizing the classical four-velocity concept.}, which generalizes the concept of velocity in stochastic mechanical formulations~\cite{Bohm1952,Nelson1966} to curved spacetime.

Substituting (\ref{eqn:11}) into (\ref{eqn:9}) and (\ref{eqn:10}), we obtain
\begin{equation}
    \label{eqn:12}
    \pi_{\mu}\pi^{\mu}+2\mathcal{A}-\frac{\hbar^2}{m^2}\frac{\Box\sqrt{n}}{\sqrt{n}}=0,
\end{equation}
\begin{equation}
    \label{eqn:13}
    \nabla^{\mu}\left(n\pi_{\mu}\right)=0,
\end{equation}

These equations constitute the geodesic formulation of the Klein-Gordon-Maxwell equation (\ref{eqn:1}). Equation (\ref{eqn:12}) resembles the quantum Hamilton-Jacobi equation, while (\ref{eqn:13}) represents the continuity equation for the bosonic system. This formulation extends previous work on boson gases in curved spacetime~\cite{Escobar2023}. We can therefore define the probability density current as
\begin{equation}
    \label{eqn:14}
    J_{\mu}:=n\pi_{\mu},
\end{equation}
which demonstrates that the probability flow follows the geodesic velocity $\pi_{\mu}$ through spacetime, consistent with the hydrodynamic interpretation of quantum mechanics~\cite{Madelung1927}.
\section{Derivation of the Balance equation}

We define the potential
\begin{equation}
U^Q = -\frac{\hbar^2}{2m^2}\frac{\square\sqrt{n}}{\sqrt{n}},
\label{eqn:15}
\end{equation}
which corresponds to the quantum potential or Bohm potential, a key concept in hydrodynamic formulations of quantum mechanics \cite{Madelung1927, Simeonov2025}.

Thus, Eq. (\ref{eqn:12}) takes the form
\begin{equation}
\pi_{\mu}\pi^{\mu} + 2\mathcal{A} + 2U^{Q} = 0.
\label{eqn:16}
\end{equation}

Applying $\nabla^{\alpha}$ to (\ref{eqn:16}) using the Leibniz rule for the first term and the Faraday tensor definition (3), we obtain
\begin{equation}
\pi_{\mu}\nabla^{\mu}\pi_{\alpha} = F_{\alpha}^{E} + F_{\alpha}^{Q} + F_{\alpha}^{n},
\label{eqn:17}
\end{equation}
where
\begin{equation}
F_{\mu}^{Q} = -\nabla_{\mu}U^{Q},
\label{eqn:18}
\end{equation}
represents the quantum force,
\begin{equation}
F_{\mu}^{n} = -\nabla_{\mu}\mathcal{A},
\label{eqn:19}
\end{equation}
the effective force due to self-interactions, and
\begin{equation}
F_{\mu}^{E} = \frac{e}{m}\pi^{\nu}F_{\nu \mu},
\label{eqn:20}
\end{equation}
which we identify as the Lorentz force per unit mass in curved spacetime, extending the classical electromagnetic force to the relativistic domain \cite{Greiner2000}.

Equation (\ref{eqn:17}) describes the dynamics of the system (equivalent to the Euler equation for quantum fluids), while Eqs. (\ref{eqn:12}) and (\ref{eqn:13}) are dynamically equivalent to the original Klein-Gordon-Maxwell equations. This force decomposition provides a clear physical interpretation of the different contributions governing the system's evolution in curved spacetime \cite{Chavanis2017_b, Matos2019}.

\subsection{Energy Balance Equation: Thermodynamical First Law}
\label{sec:energy_balance}

To derive our energy balance equation, we contract the force equation (\ref{eqn:17}) with $n\nabla^{\alpha}t$, where $t$ is the time function of the ADM foliation. This approach differs from that of Matos et al. ~\cite{Matos2019}, who contracted with $nv^{\mu}$, with $v_{\mu}$ being a hydrodynamic velocity related to $\pi_{\mu}$ by
\begin{equation}
\pi_{\mu} = v_{\mu} + \frac{\hbar\omega_{0}}{m}\nabla_{\mu}t,
\label{eqn:21}
\end{equation}
where $\omega_0$ is the frequency of the state. Here $v_{\mu}$ represents the velocity associated with the flow of the boson fluid within the hydrodynamic framework, analogous to classical fluid velocity but in a relativistic quantum context \cite{Chavanis2017_b}. Alternatively, contracting (\ref{eqn:17}) with $n\nabla^{\alpha}t$, we obtain
\begin{equation}
(n\nabla^{\alpha}t)\pi_{\mu}\nabla^{\mu}\pi_{\alpha} = n\nabla^{\alpha}t\left(F_{\alpha}^{E} + F_{\alpha}^{Q} + F_{\alpha}^{n}\right).
\label{eqn:22}
\end{equation}

Using the continuity equation (\ref{eqn:13}), the left side of Eq. (\ref{eqn:22}) can be rewritten as
\begin{equation}
(n\nabla^{\alpha}t)\pi_{\mu}\nabla^{\mu}\pi_{\alpha} = n\pi_{\mu}g^{\alpha 0}\nabla^{\mu}\pi_{\alpha}
= \nabla^{\mu}(n\pi_{\mu}\pi^{0}),
\label{eqn:23}
\end{equation}
while on the right side, we obtain
\begin{equation}
n(\nabla^{\alpha}t)F_{\alpha}^{E} = nF^{E0}
= \frac{ne}{m}\pi_{\mu}F^{\mu 0} = -\frac{1}{m}\nabla^{\mu}\mathcal{P}_{\mu},
\label{eqn:24}
\end{equation}
where $\mathcal{P}_{\mu}$ is the Poynting 4-vector, representing electromagnetic energy flux in curved spacetime \cite{Greiner2000}. For $F^Q$, we get
\begin{equation}
n(\nabla^{\alpha}t)F_{\alpha}^{Q} = nF^{Q0}
= -n\nabla^{0}U^{Q} = -\nabla^{\mu}J_{\mu}^{Q},
\label{eqn:25}
\end{equation}
where we define the quantum 4-current as
\begin{equation}
J_{\mu}^{Q} = -\frac{\hbar^{2}}{m^{2}} n\nabla^{0}\nabla_{\mu}(\ln n).
\label{eqn:26}
\end{equation}

Combining results (\ref{eqn:23}), (\ref{eqn:24}), and (\ref{eqn:25}) with (\ref{eqn:19}), Eq. (\ref{eqn:22}) can be rewritten as
\begin{equation}
\nabla^{\mu}(J_{\mu}\pi^{0}) + \frac{1}{m}\nabla^{\mu}\mathcal{P}_{\mu} + \nabla^{\mu}J_{\mu}^{Q} + n\nabla^{0}\mathcal{A} = 0.
\label{eqn:27}
\end{equation}

Defining the total energy flux
\begin{equation}
\mathcal{J}^{\mu} := J^{\mu}\pi^{0} + \frac{1}{m}\mathcal{P}^{\mu} + J^{Q\mu},
\label{eqn:28}
\end{equation}
we obtain the energy balance equation
\begin{equation}
\nabla^{\mu}\mathcal{J}_{\mu} + n\nabla^{0}\mathcal{A} = 0.
\label{eqn:29}
\end{equation}

This equation reveals the relationship between energy flux and changes in the effective potential $\mathcal{A}$ acting as sources and sinks for $\mathcal{J}_{\mu}$. This formulation is analogous to the first law of thermodynamics and represents the energy balance equation for the boson gas system in a spacetime of arbitrary curvature, extending previous thermodynamic descriptions of relativistic quantum fluids \cite{Matos2019, Jimnez-Aquino2018}.

\subsection{Integration over Spatial Hypersurface $\Sigma_{t}$}
\label{sec:integration}

To obtain a global conservation law, we integrate the energy balance equation (\ref{eqn:29}) over a spatial hypersurface $\Sigma_t$. Using the ADM metric form given in Eq. (5), the volume element is $\sqrt{-g} = N\sqrt{\gamma}$. The integration yields:
\begin{equation}
\begin{split}
&\int_{\Sigma_{t}}\nabla_{\mu}\mathcal{J}^{\mu}\sqrt{-g} d^{3}x
+\\
&\int_{\Sigma_{t}}n\left(-\frac{1}{N^{2}}\partial_{0}\mathcal{A} + \frac{N^{i}}{N^{2}}\partial_{i}\mathcal{A}\right)\sqrt{-g} d^{3}x = 0.
\end{split}
\label{eqn:30}
\end{equation}

Expanding the covariant divergence into partial derivatives, we obtain
\begin{equation}
\begin{split}
&\int_{\Sigma_{t}}\partial_{0}(N\sqrt{\gamma}\mathcal{J}^{0})d^{3}x + \int_{\Sigma_{t}}\partial_{i}(N\sqrt{\gamma}\mathcal{J}^{i})d^{3}x\\
&-\int_{\Sigma_{t}}\frac{n\sqrt{\gamma}}{N}\partial_{0}\mathcal{A}d^{3}x + \int_{\Sigma_{t}}\frac{n\sqrt{\gamma}N^{i}}{N}\partial_{i}\mathcal{A}d^{3}x = 0.
\end{split}
\label{eqn:31}
\end{equation}

We now separate the temporal\footnote{For simplicity, we will now refer to the evolution variable $t$ as time, although it is not actually the system time, just a parameter of the evolution.} and spatial contributions. Applying the divergence theorem to the spatial integral and identifying the time derivative of the integrated energy, we find:
\begin{equation}
\begin{split}
\frac{d}{dt}\int_{\Sigma_t}N\sqrt{\gamma}\mathcal{J}^0 d^3 x + \oint_{\partial \Sigma_t}Nc\sqrt{\gamma}\mathcal{J}^i dS_i\\
-\int_{\Sigma_t}\frac{n\sqrt{\gamma}}{N}\partial_t\mathcal{A}d^3 x + \oint_{\partial \Sigma_t}\frac{nc\sqrt{\gamma}N^i}{N}\mathcal{A}dS_i = 0.
\end{split}
\label{eqn:32}
\end{equation}

The third term, which involves the time derivative of the interaction potential $\mathcal{A}$, requires careful treatment. It can be rewritten as
\begin{equation}
\begin{split}
\int_{\Sigma_t}\frac{n\sqrt{\gamma}}{N}\partial_t\mathcal{A}d^3 x = & \frac{d}{dt}\int_{\Sigma_t}\frac{n\sqrt{\gamma}}{N}\mathcal{A}d^3 x\\
&-\int_{\Sigma_t}\mathcal{A}\partial_t\left(\frac{n\sqrt{\gamma}}{N}\right)d^3 x.
\end{split}
\label{eqn:33}
\end{equation}

The last integral in Eq. (\ref{eqn:33}) is not a total time derivative; it represents an exchange of energy between the matter distribution and the evolving spacetime geometry. Expanding it further, we obtain
\begin{equation}
\begin{split}
\int_{\Sigma_t}\mathcal{A}\partial_t\left(\frac{n\sqrt{\gamma}}{N}\right)d^3 x = & \int_{\Sigma_t}\mathcal{A}\frac{\sqrt{\gamma}}{N}\partial_tn d^3 x\\
&+\int_{\Sigma_t}n\mathcal{A}\partial_t\left(\frac{\sqrt{\gamma}}{N}\right)d^3 x.
\end{split}
\label{eqn:34}
\end{equation}

Collecting all terms, we arrive at the following global balance equation:
\begin{equation}
\frac{d}{dt}\left[\int_{\Sigma_t}\mathcal{E}d^3 x + W\right] + \mathrm{Flux} + \Phi_I - \mu_s + \mathcal{T} = 0,
\label{eqn:35}
\end{equation}
In this expression, the local energy density is defined as $\mathcal{E} = N\sqrt{\gamma}\mathcal{J}^0$, the interaction energy as $W = -\int_{\Sigma_t} (n\sqrt{\gamma}/N)\mathcal{A} d^3x$, the boundary energy flux as $\mathrm{Flux} = \oint_{\partial \Sigma_t} Nc\sqrt{\gamma}\mathcal{J}^i dS_i$, the electromagnetic contribution as $\Phi_I = \oint_{\partial \Sigma_t} (nc\sqrt{\gamma}N^i/N)\mathcal{A} dS_i$, and the chemical potential term as $\mu_s = \int_{\Sigma_t} \mathcal{A} (\sqrt{\gamma}/N) \partial_t n d^3x$. The term $\mathcal{T}$ is defined in Eq. (\ref{eqn:37}),
where $\mathcal{E}$ represents the local energy density, $W$ accounts for interaction energy, $\mathrm{Flux}$ denotes boundary energy flows, $\Phi_I$ captures electromagnetic contributions, $\mu_s$ is the chemical potential associated with particle number changes, and $\mathcal{T}$ encodes the energy exchange with the dynamical spacetime geometry.

In the first term on the right side, we identify the chemical potential in curved spacetime. This represents energy transfer due to changes in boson number density $n$, following thermodynamic descriptions of relativistic quantum systems \cite{Matos2019}. In collapsing systems ($\partial_t n > 0$), this term acts as an energy sink, converting kinetic energy into interaction energy. The ADM formalism employed here provides a natural framework for such thermodynamic analyses in dynamical spacetimes \cite{Alcubierre2008}.

We can rewrite Eq. (\ref{eqn:35}) in a more familiar thermodynamic form:
\begin{equation}
\frac{dU}{dt} +\mathrm{Flux} + \Phi_I - \mu_s + \mathcal{T} = 0,
\label{eqn:36}
\end{equation}
where $U$ represents the total internal energy including interaction effects. We thus recover the first law of thermodynamics in curved spacetime, and Eq. (\ref{eqn:36}) provides an improved formulation of the energy balance equation that explicitly accounts for gravitational and quantum contributions. This formulation extends previous thermodynamic descriptions of bosonic systems \cite{Matos2019} while incorporating the geometric structure of spacetime through the ADM formalism \cite{Alcubierre2008}. The resulting framework offers a more complete thermodynamic description for relativistic quantum gases in dynamical spacetimes.

\subsection{First Law of Thermodynamics Form and Thermodynamical Spacetime Coupling $\mathcal{T}$}
\label{sec:first_law}

The global balance equation (\ref{eqn:35}) can be recast in a form that resembles the first law of thermodynamics in curved spacetime. Defining the total internal energy $U = \int_{\Sigma_t} \mathcal{E} d^3 x + W$, we obtain Eq. (\ref{eqn:36}), which expresses the conservation of energy including gravitational and quantum contributions.

The thermodynamic coupling term $\mathcal{T}$ captures the dynamic energy exchange between spacetime geometry and matter distribution:
\begin{equation}
\mathcal{T} = -\int_{\Sigma_t} \mathcal{A} \partial_t\left(\frac{\sqrt{\gamma}}{N}\right) d^3 x,
\label{eqn:37}
\end{equation}
$\mathcal{T}$ decomposes into two fundamental physical contributions. The first, spatial volume dynamics, encodes energy exchange from the expansion or contraction of spatial hypersurfaces. For $\partial_t\sqrt{\gamma} >0$ (expansion), energy is transferred from matter to geometry, generalizing $PdV$ work to dynamical spacetime, following the thermodynamic approach to gravitational systems [24]. The second contribution, lapse function variations, accounts for energy changes due to gravitational time dilation effects. Temporal variations in the lapse function $N$ redistribute energy between matter and the gravitational field, reflecting the role of the lapse function as a gravitational potential in the ADM formalism \cite{Alcubierre2008}.

\paragraph{Physical Interpretation:}
\begin{itemize}
\item $\mathcal{T} < 0$: Energy flows from matter to spacetime geometry,
\item $\mathcal{T} > 0$: Energy transfers from geometry to matter fields.
\item In stationary spacetimes: $\partial_t(\cdot) = 0 \Rightarrow \mathcal{T} = 0$
\end{itemize}

This decomposition provides a clear thermodynamic interpretation of the energy transfer mechanisms in relativistic bosonic systems, extending previous work on gravitational thermodynamics \cite{Hawking1975} and offering new insights into energy exchange processes in dynamical spacetimes \cite{Matos2019}.

\section{Stochastic velocity: $u_{\mu}$}

A stochastic component in the spacetime background can be incorporated
into the description of particle motion without invoking quantum gravity.
In the present framework, this is implemented through the stochastic
velocity $u_{\mu}$, which accounts for the random diffusive contribution
to the particle's trajectory induced by a fluctuating background. The
resulting dynamics can be cast in the form of a generalized Langevin
equation on a curved manifold, where the stochastic term models the
effect of background fluctuations and the dissipative term ensures
consistency with the fluctuation-dissipation structure of the theory
\cite{Nelson1966, Jimnez-Aquino2018}. The process is assumed to be
Markovian, so that the evolution depends only on the current state.

Mathematically, the stochastic velocity is defined as
$u_{\mu} = (\hbar/2m)\nabla_{\mu}\ln n$, and together with the geodesic
velocity $\pi_{\mu}$ it forms the complete complex velocity
$\eta_{\mu} = \pi_{\mu} - i u_{\mu}$ that characterizes the flow of the
boson gas. The introduction of this stochastic component is a technical
device that allows us to separate the diffusive and conservative aspects
of the dynamics, and it is this separation that ultimately leads to the
Fisher information constraint derived in Section~6. No assumption is made
about the fundamental origin of the stochastic background---it may arise
from a classical stochastic background of gravitational waves
\cite{Hu2008, Escobar2023, Carney2024}, from quantum fluctuations of the metric,
or from other sources. The only requirement is that the background
fluctuations can be treated statistically and that the Markovian
approximation holds. The central goal of this manuscript is not to
identify the origin of such fluctuations, but to study their
thermodynamic and information-theoretic consequences within the
Madelung-ADM hydrodynamic framework on a fixed curved spacetime
\cite{Brown1992, Escobar-Aguilar2025, delaPea2015}.

We define the following differential operators for curved spacetime:
\begin{equation}
D_{c} = \pi^{\mu}\nabla_{\mu},
\label{eqn:38}
\end{equation}
\begin{equation}
D_{s} = u^{\mu}\nabla_{\mu} + \lambda \square,
\label{eqn:39}
\end{equation}
where $D_{c}$ and $D_{s}$ are systematic derivative and stochastic derivative respectively, $\lambda$ the diffusion coefficient, in this quantum context must be $\lambda = \hbar /2m$, following Nelson's stochastic quantization approach \cite{Nelson1966}. And $u_{\mu}$ \footnote{Other authors call it the ``osmotic velocity'' \cite{Nelson1966} or the ``quantum velocity'' \cite{Chavanis2017, Chavanis2024}.} is defined by
\begin{equation}
u_{\mu} = \frac{\hbar}{2m}\nabla_{\mu}\ln (n).
\label{eqn:40}
\end{equation}
Within this framework, we define the complete derivative $D$ as\footnote{The parameter $\iota = i$ in the complete derivative $D = D_c + \iota D_s$ ensures compatibility with the Schr\"{o}dinger equation and introduces the necessary complex structure for quantum interference \cite{Nelson2012, Beyer2021, Escobar2023}.}
\begin{equation}
D = D_{c} + \iota D_{s},
\label{eqn:41}
\end{equation}
 This definition completes derivative $D$
with $\iota$ an arbitrary parameter. From Eqs. (\ref{eqn:6}), (\ref{eqn:11}), (\ref{eqn:21}) and consistency with the Schr\"{o}dinger equation, the optimal choice is $\iota = -i$. This complex structure emerges naturally in stochastic quantum mechanics \cite{Nelson2012, Beyer2021, Escobar-Aguilar2025}.

We note the following relations
\begin{equation}
Dx_{\mu} = \pi_{\mu} - iu_{\mu} = \eta_{\mu},
\label{eqn:42}
\end{equation}
where $\eta_{\mu}$ is the complete velocity~\cite{delaPea2015, Chavanis2017_b, Chavanis2024}  and is the trajectory of the boson. In other words, (\ref{eqn:42}) provides a method to calculate trajectories in the presence of gravitational wave background through stochastic considerations.
\begin{equation}
D_{c}x_{\mu} = \pi_{\mu},
\label{eqn:43}
\end{equation}
\begin{equation}
D_{s}x_{\mu} = u_{\mu}.
\label{eqn:44}
\end{equation}

These decomposition relations reveal the fundamental structure of the stochastic dynamics: Equation (\ref{eqn:43}) shows that the systematic derivative $D_{c}$ acting on the position yields the geodesic velocity $\pi_{\mu}$, which despite its classical appearance contains quantum information through the phase $\theta$ in its definition (\ref{eqn:11}). This represents the quantum-corrected deterministic component of the motion. Conversely, Equation (\ref{eqn:44}) demonstrates that the stochastic derivative $D_{s}$ acting on position produces the purely stochastic velocity $u_{\mu}$, capturing the random, diffusive component arising from spacetime fluctuations. The complex combination in (\ref{eqn:42}) thus represents the complete quantum-stochastic velocity, where both components ($\pi_{\mu}$ and $u_{\mu}$) are inherently quantum in nature, with $\pi_{\mu}$ encoding phase information and $u_{\mu}$ encoding density gradient information from the quantum probability distribution.

\section{Fisher Entropy}
\label{sec:fisher_entropy}
In information theory, the Fisher information quantifies the amount of information that an observable random variable carries about an unknown parameter upon which the probability distribution depends~\cite{Frieden2004}. Within the Madelung hydrodynamic formulation of quantum mechanics, it provides a natural measure of ``quantum uncertainty'' or ``structural information'' embedded in the wavefunction's spatial variations~\cite{Madelung1927, Simeonov2025}. When applied to spacetime coordinates in a curved background, it connects quantum dynamics with the geometry itself, offering an information-theoretic constraint on the system's evolution~\cite{Escobar2023, Chavanis2024}. We now derive its covariant expression and explore its physical interpretation.

The Fisher entropy \(I_{F}\) quantifies the information content of a probability distribution---specifically, how much a single measurement reduces our uncertainty about the system~\cite{Frieden2004, DnesPetz2008}. In the context of the Madelung hydrodynamic formulation of quantum mechanics~\cite{Madelung1927, Simeonov2025}, the Fisher entropy corresponds to the kinetic term in the Lagrangian density, minus a non-local term that ensures gauge invariance.

\subsection{Fisher Entropy in Curved Spacetime}
\label{sec:fisher_curved}

The quantum Fisher information for estimating a spacetime coordinate parameter $x^{\mu}$ is defined via the symmetric logarithmic derivative (SLD) operator $L_{\mu}$\footnote{The symmetric logarithmic derivative operator $L_{\mu}$ is defined implicitly by the relation $\frac{d\rho}{dx^{\mu}} = \frac{1}{2}(\rho L_{\mu} + L_{\mu}\rho)$ and provides the optimal estimator for parameter estimation in quantum systems \cite{DnesPetz2008}} as
\begin{equation}
I_{F}(x^{\mu}) = \mathrm{Tr}[\rho L_{\mu}^{2}],
\label{eqn:45}
\end{equation}
where the SLD operator $L_{\mu}$ is implicitly defined by the relation
\begin{equation}
\frac{d\rho}{dx^{\mu}} = \frac{1}{2} (\rho L_{\mu} + L_{\mu}\rho).
\label{eqn:46}
\end{equation}

Here $\rho = |\Phi \rangle \langle \Phi |$ is the pure state density matrix of the boson gas. For a non-normalized state, we introduce the total boson number
\begin{equation}
\mathcal{N} = \langle \Phi |\Phi \rangle = \int_{\Sigma_t}d^3 x\sqrt{-g} n,
\label{eqn:47}
\end{equation}
which allows us to write the SLD explicitly as
\begin{equation}
\begin{split}
L_{\mu} = & \frac{2}{\mathcal{N}} \left( |\partial_{\mu}\Phi \rangle \langle \Phi | + |\Phi \rangle \langle \partial_{\mu}\Phi | \right)\\
&- \frac{\langle\Phi|\partial_{\mu}\Phi\rangle + \langle\partial_{\mu}\Phi|\Phi\rangle}{\mathcal{N}^2} |\Phi \rangle \langle \Phi |.
\end{split}
\label{eqn:48}
\end{equation}

Substituting the Madelung representation $\Phi = \sqrt{n} e^{i\theta}$ and the covariant derivative $D_{E\mu} = \nabla_{\mu} + i\frac{e}{\hbar} A_{\mu}$ into the definition of the Fisher information yields
\begin{equation}
I_{F}(x^{\mu}) = 4\langle D_{E\mu}\Phi |D_{E\mu}\Phi \rangle -\frac{4[\mathrm{Im}\langle\Phi |D_{E\mu}\Phi \rangle ]^{2}}{\mathcal{N}}.
\label{eqn:49}
\end{equation}

The first term captures the total sensitivity of the state to local displacements, while the second term subtracts the contribution arising from a global phase shift, thereby enforcing the $U(1)$ gauge invariance of the Fisher information. The quantum Fisher information density is proportional to $\langle D_{E\mu}\Phi |D_{E\mu}\Phi \rangle$ which corresponds to the kinetic term in the Lagrangian density, minus a non-local term that ensures gauge invariance.

\subsection{Hydrodynamic Representation}
\label{sec:hydrodynamic_representation}

Using the definitions of the stochastic velocity $u_{\mu} = \frac{\hbar}{2m}\nabla_{\mu}\ln n$ and the geodesic velocity $\pi_{\mu} = \frac{\hbar}{m}\big(\nabla_{\mu}\theta +\frac{e}{\hbar} A_{\mu}\big)$, both having dimensions of velocity, we evaluate the two expectation values appearing in Eq. (\ref{eqn:49}). First,
\begin{equation}
\begin{split}
\langle D_{E\mu}\Phi |D_{E\mu}\Phi \rangle = & \int_{\Sigma_t}d^3 x\sqrt{-g} |D_{E\mu}\Phi |^2\\
= & \frac{m^2}{\hbar^2}\int_{\Sigma_t}d^3 x\sqrt{-g} n\big(u^\mu u_\mu +\pi^\mu\pi_\mu\big).
\end{split}
\label{eqn:50}
\end{equation}

Second, the imaginary part of the mixed expectation value is
\begin{equation}
\mathrm{Im}\langle \Phi |D_{E\mu}\Phi \rangle = \frac{m}{\hbar}\int_{\Sigma_t}d^3 x\sqrt{-g} n\pi_{\mu}.
\label{eqn:51}
\end{equation}

Inserting Eqs. (\ref{eqn:50}) and (\ref{eqn:51}) into Eq. (\ref{eqn:49}) gives the hydrodynamic expression of the Fisher information:
\begin{equation}
\begin{split}
I_{F}(x^{\mu}) = & \frac{4m^{2}}{\hbar^{2}}\int_{\Sigma_{t}}d^{3}x\sqrt{-g} n\big(u^\mu u_\mu +\pi^\mu\pi_\mu\big)\\
&-\frac{4m^{2}}{\hbar^{2}\mathcal{N}}\Big|\int_{\Sigma_{t}}d^{3}x\sqrt{-g} n\pi_{\mu}\Big|^{2}.
\end{split}
\label{eqn:52}
\end{equation}

The second term in Eq. (\ref{eqn:52}) is non-local and removes the contribution proportional to the square of the integrated current $\int n\pi_{\mu}\sqrt{- g} d^{3}x$. Here $\pi_{\mu} = \frac{\hbar}{m} (\nabla_{\mu}\theta +\frac{e}{\hbar} A_{\mu})$ is the geodesic velocity, which combines phase gradients and electromagnetic coupling. While the continuity equation $\nabla^{\mu}(n\pi_{\mu}) = 0$ guarantees conservation of the total charge $Q = \int n\pi^{0}\sqrt{- g} d^{3}x$, it does not force the spatial components $\int n\pi_{i}\sqrt{- g} d^{3}x$ to vanish. Consequently, the non-local term persists in general dynamical settings, ensuring that $I_{F}$ quantifies only the local structural information contained in density and phase inhomogeneities, not the global current or overall phase.

Introducing the complete quantum-stochastic velocity~\cite{Chavanis2017,Chavanis2024} $\eta_{\mu} = \pi_{\mu} - iu_{\mu}$ from Eq. (\ref{eqn:42}), we note that
\begin{equation}
|\eta_{\mu}|^{2} = u^\mu u_\mu +\pi^\mu\pi_\mu,
\label{eqn:53}
\end{equation}
which allows us to rewrite the first term of Eq. (\ref{eqn:52}) in the compact form
\begin{equation}
\int_{\Sigma_t}d^3 x\sqrt{-g} n\big(u^\mu u_\mu +\pi^\mu\pi_\mu\big) = \int_{\Sigma_t}d^3 x\sqrt{-g} n|\eta_\mu |^2.
\label{eqn:54}
\end{equation}

Equation (\ref{eqn:54}) highlights that the local Fisher information density is proportional to the squared magnitude of the complex velocity $\eta_{\mu}$, which unifies the deterministic (geodesic) and stochastic (osmotic) components of the quantum flow.

\subsection{Continuity equation and the integrated current}
The continuity equation $\nabla^{\mu}(n\pi_{\mu}) = 0$ holds as a local conservation law. Integrating it over a spacetime volume $\nu$ bounded by two spacelike hypersurfaces $\Sigma_{t_1}$ and $\Sigma_{t_2}$ and timelike boundaries at spatial infinity, and applying the Gauss theorem in curved spacetime, we obtain
\begin{equation}
\int_{\nu}\nabla^{\mu}(n\pi_{\mu})\sqrt{-g} d^{4}x = \oint_{\partial \nu}n\pi^{\mu}d\Sigma_{\mu} = 0,
\label{eqn:55}
\end{equation}
where $d\Sigma_{\mu}$ is the outward-pointing volume element on the boundary. Equation (\ref{eqn:55}) is an identity following from the continuity equation $\nabla^{\mu}(n\pi_{\mu}) = 0$; it does not, by itself, force the integrated current $\int n\pi_{\mu}\sqrt{- g} d^{3}x$ to vanish.

To determine when the non-local term in Eq. (\ref{eqn:52}) disappears, we must examine the integrated current vector
\begin{equation}
J_{\mu}^{\mathrm{int}}(t)\equiv \int_{\Sigma_{t}}n\pi_{\mu}\sqrt{-g} d^{3}x.
\label{eqn:56}
\end{equation}

The temporal component $J_{0}^{\mathrm{int}}$ is conserved, $\partial_{t}J_{0}^{\mathrm{int}} = 0$, as follows from Eq. (\ref{eqn:55}) with suitable boundary conditions. The spatial components $J_{i}^{\mathrm{int}}$, however, are not constrained by conservation laws alone.

The condition for the non-local term to vanish is therefore
\begin{equation}
J_{\mu}^{\mathrm{int}}(t) = 0 \quad \mathrm{for~the~parameter~direction~}\mu.
\label{eqn:57}
\end{equation}

This occurs in several physically relevant regimes which can be derived from the structure of Eq. (\ref{eqn:55}) together with additional symmetry or stationarity assumptions. For instance, in stationary spacetimes with static matter, where the metric and density $n$ are time-independent and the phase satisfies $\partial_t\theta =$ const., the spatial current is divergence-free, and if the flux through the boundary vanishes, then $J_{i}^{\mathrm{int}} = 0$. Similarly, if the spatial hypersurface $\Sigma_{t}$ is maximally symmetric (e.g., Euclidean $\mathbb{R}^{3}$ or a 3-sphere) and the configuration is isotropic, then $n\pi_{i}$ must be proportional to a Killing vector, and the only isotropic vector field on such a space is the zero field, forcing $J_{i}^{\mathrm{int}} = 0$. If the hypersurface $\Sigma_{t}$ admits a reflection isometry under which $n$ is even and $\pi_{i}$ is odd, then the integrand $n\pi_{i}$ is odd and integrates to zero. Finally, the current $n\pi_{\mu}$ shifts under a gauge transformation $\theta \rightarrow \theta +\alpha(x)$, $A_{\mu} \rightarrow A_{\mu} - \partial_{\mu}\alpha$: choosing $\alpha$ such that $\int n(\nabla_{\mu}\theta +\frac{\epsilon}{\hbar} A_{\mu})\sqrt{-g}d^{3}x = 0$ sets $J_{\mu}^{\mathrm{int}} = 0$ without altering physical observables.

When any of the above conditions hold, $J_{\mu}^{\mathrm{int}} = 0$ and the non-local term in Eq. (\ref{eqn:52}) vanishes, reducing the Fisher information to the integrated local density
\begin{equation}
I_{F}(x^{\mu}) = \frac{4m^{2}}{\hbar^{2}}\int_{\Sigma_{t}}d^{3}x\sqrt{-g} n|\eta_{\mu}|^{2}.
\label{eqn:58}
\end{equation}

In generic dynamical spacetimes, however, $J_{\mu}^{\mathrm{int}}$ need not be zero, and the full expression (\ref{eqn:52}) must be retained. The non-local term therefore measures the departure from the symmetry or stationarity conditions that decouple global current contributions from local information content.

\subsection{Fisher entropy density} We define the Fisher entropy density as the integrand of the local approximation:
\begin{equation}
\mathcal{I}_F(x) = \frac{4m^2}{\hbar^2} n|\eta_\mu |^2 = \frac{4m^2}{\hbar^2} n(u_\mu^2 +\pi_\mu^2).
\label{eqn:59}
\end{equation}

Using the relation $\pi_{\mu}\pi^{\mu} + 2\mathcal{A} + 2U^{Q} = 0$ from Eq. (\ref{eqn:16}) and the expression of the quantum potential $U^{Q} = - \frac{\hbar^{2}}{2m^{2}}\frac{\square\sqrt{n}}{\sqrt{n}}$, we can rewrite $\mathcal{I}_{F}$ in a form that highlights its connection to the dynamics of the boson density:
\begin{equation}
\mathcal{I}_F = 2\square n - \frac{8m^2}{\hbar^2} n\mathcal{A}.
\label{eqn:60}
\end{equation}

Equation (\ref{eqn:60}) shows that the Fisher entropy density is sourced by two competing effects: the spacetime Laplacian of the density $\square n$, which encodes quantum-stochastic diffusion, and the self-interaction potential $\mathcal{A}$, which tends to localize the boson cloud.

\subsection{Information-theoretic constraint} Combining the full expression for $I_{F}$ from Eq. (\ref{eqn:52}) with the local density form (\ref{eqn:60}), we obtain the information-theoretic constraint for the boson gas in curved spacetime:
\begin{equation}
\begin{split}
I_F(x^{\mu})
+\frac{4m^2}{\hbar^2\mathcal{N}}\Big|\int_{\Sigma_t}d^3 x\sqrt{-g} n\pi_{\mu}\Big|^2=\\
\int_{\Sigma_t}d^3 x\sqrt{-g}\left[2\square n -\frac{8m^2}{\hbar^2} n\mathcal{A}\right] .
\end{split}
\label{eqn:61}
\end{equation}

This constraint separates the locally measurable information (right-hand side) from the global phase contribution (left-hand side). In the absence of a net integrated current, it reduces to a local conservation law for quantum information.

\subsection{Physical interpretation} The Fisher entropy density $\mathcal{I}_F(x)$ quantifies the structural information stored in the boson cloud: regions of high density gradient or rapid phase variation correspond to high Fisher entropy. The non-local term in Eq. (\ref{eqn:52}) acts as an information regulator that discards gauge-dependent global phase data, ensuring that $I_{F}$ remains an observable measure of quantum uncertainty.

In stationary configurations (e.g., boson stars, atomic orbitals), the integrated current vanishes and the Fisher information coincides with the integral of $\mathcal{I}_F$, providing a direct link between information content and hydrodynamic energy densities. In dynamical spacetimes, the non-local term must be retained, reflecting the interplay between quantum information and global causal structure.

The complete description given by Eqs. (\ref{eqn:45})-(\ref{eqn:61}) unifies the thermodynamics of relativistic boson gases with information-theoretic principles, offering a refined foundation for studying systems where quantum, gravitational, and informational aspects are inseparably intertwined \cite{Frieden2004, DnesPetz2008, Matos2019, Chavanis2024}.

\section{Examples}
\label{sec:examples}

In this section, we illustrate the applicability and consistency of our formalism in three representative systems: the harmonic oscillator and hydrogen atom in Minkowski spacetime, and the Klein--Gordon field in Schwarzschild geometry. These examples serve to explicitly verify the fundamental equations of our theoretical framework---the energy balance equation (\ref{eqn:29}) and the information-entropy constraint (\ref{eqn:60})---while simultaneously revealing the role of the stochastic velocity $u_\mu$ and the dynamics of quantum information in different relativistic contexts.

\subsection{Minkowski Spacetime}
\label{subsec:minkowski}

In the limit of flat spacetime, described by the Minkowski metric
\begin{equation}
ds^{2} = -c^{2}dt^{2} + dx_{i}dx^{i},
\label{eqn:62}
\end{equation}
our covariant formalism simplifies considerably, allowing for explicit analytical solutions. We consider two canonical quantum systems---the harmonic oscillator and the hydrogen atom---to validate the internal consistency of the model and connect our results with non-relativistic quantum mechanics.

\subsubsection{One-dimensional Harmonic Oscillator}
\label{subsubsec:harmonic_oscillator}

For the harmonic oscillator, the self-interaction potential in our formalism is \cite{Gallegos2021}
\begin{equation}
\mathcal{A}(x) = \frac{\omega^{2}}{2} x^{2},
\label{eqn:63}
\end{equation}
which is time-independent. The Klein--Gordon equation (\ref{eqn:1}) admits separation of variables, $\Phi(x_{\mu}) = \Psi(\vec{x})\phi(t)$, with the temporal part given by
\begin{equation}
\phi(t) = A e^{i\frac{E}{\hbar} t} + B e^{-i\frac{E}{\hbar} t},
\label{eqn:64}
\end{equation}
and the spatial part satisfying
\begin{equation}
\nabla^2\Psi + \left( \frac{E^2}{\hbar^2 c^2} - \frac{2m^2}{\hbar^2} \mathcal{A} \right) \Psi = 0.
\label{eqn:65}
\end{equation}
The normalized solution corresponds to the well-known stationary states of the quantum harmonic oscillator:
\begin{equation}
\Psi_{\nu}(x) = \sqrt{\frac{1}{2^{\nu}\nu!}}\left(\frac{m\omega}{\pi\hbar}\right)^{1/4} H_{\nu}\!\left(\sqrt{\frac{m\omega}{\hbar}} x\right) e^{-\frac{m\omega x^{2}}{2\hbar}},
\label{eqn:66}
\end{equation}
where $H_\nu$ are the Hermite polynomials, with quantized energy levels
\begin{equation}
E^2 = \hbar \omega m c^2 (2\nu + 1),\quad \nu+0,1,2\dotsc
\label{eqn:67}
\end{equation}

The probability density is then
\begin{equation}
\begin{split}
n = & \\ 
|\Phi_{\nu}|^2 = & \frac{1}{2^{\nu}\nu!} \left( \frac{m\omega}{\pi\hbar} \right)^{1/2} H_{\nu}^2\left( \sqrt{\frac{m\omega}{\hbar}} x \right) e^{-\frac{m\omega x^2}{\hbar}} F(t),
\end{split}
\label{eqn:68}
\end{equation}
where $F(t) = |\phi(t)|^{2}$.

The phase $\theta$ follows from $\Phi = \sqrt{n} e^{i\theta}$ as
\begin{equation}
\theta = -i \ln \left[ \frac{A e^{i\frac{E}{\hbar} t} + B e^{-i\frac{E}{\hbar} t}}{\sqrt{F(t)}} \right].
\label{eqn:69}
\end{equation}

The geodesic velocity components, from definition (\ref{eqn:11}), are
\begin{equation}
\pi_0 = \frac{E}{\hbar c} \frac{|A|^2 - |B|^2}{F(t)}, \qquad \pi_x = 0.
\label{eqn:70}
\end{equation}

The quantum potential (\ref{eqn:15}) becomes
\begin{equation}
U^{Q} = \frac{E^2}{2m^2c^4}\frac{(|A|^{2} - |B|^{2})^{2}}{F^{2}(t)} - \mathcal{A}.
\label{eqn:71}
\end{equation}

From the definition of the quantum current divergence (\ref{eqn:25}), we compute
\begin{equation}
\nabla_{\mu}J^{Q\mu} = \frac{2E}{\hbar c} |\Psi_{\nu}(x)|^{2} \frac{(|A|^{2} - |B|^{2})^{2} \dot{F}}{F^{2}(t)},
\label{eqn:72}
\end{equation}
while the divergence of the classical energy current satisfies
\begin{equation}
\nabla_{\mu}(J^{\mu}\pi^{0}) = -\frac{2E}{\hbar c} |\Psi_{\nu}(x)|^{2} \frac{(|A|^{2} - |B|^{2})^{2} \dot{F}}{F^{2}(t)}.
\label{eqn:73}
\end{equation}
The sum of these contributions, together with $n\nabla^{0}\mathcal{A}=0$ (since $\mathcal{A}$ is time-independent), verifies the energy balance equation (\ref{eqn:29}) identically.

The stochastic velocity components, from (\ref{eqn:40}), are
\begin{equation}
u_{0} = \frac{\hbar}{2m}\frac{\dot{F}(t)}{F(t)},\quad
u_{x} = \sqrt{\frac{\hbar\omega}{m}}\left[\frac{H_{\nu}'(\xi)}{H_{\nu}(\xi)} - \xi\right],
\label{eqn:74}
\end{equation}
where $\xi = \sqrt{\frac{m\omega}{\hbar}} x$.

The complete quantum-stochastic velocity $\eta_{\mu} = \pi_{\mu} - iu_{\mu}$ then reads
\begin{equation}
\begin{split}
\eta_0 = & \frac{E}{m c} \frac{|A|^2 - |B|^2}{F(t)} - i \frac{\hbar}{2m c} \frac{\dot{F}(t)}{F(t)},\\
\eta_x = & -i \sqrt{\frac{\hbar \omega}{m}} \left( \frac{H_{\nu}'(\xi)}{H_{\nu}(\xi)} - \xi \right).
\end{split}
\label{eqn:75}
\end{equation}

The chemical potential, obtained by direct calculation, is
\begin{equation}
\mu_s = \frac{E^2}{4 m^2 c^2} \dot{F}(t),
\label{eqn:76}
\end{equation}
oscillating with frequency $\Omega_{\mu} = 2 \sqrt{\frac{m \omega c^2 (2\nu + 1)}{\hbar}}$.

The Fisher entropy density, from (\ref{eqn:60}), takes the explicit form
\begin{equation}
\begin{split}
\mathcal{I}_F = & n \Bigg[ \frac{4m \omega}{\hbar} \left( \frac{H_{\nu}'(\xi)}{H_{\nu}(\xi)} - \xi \right)^2 - \frac{1}{c^4} \frac{\dot{F}^2(t)}{F^2(t)} \\
&- \frac{4E^2}{\hbar^2 c^4} \frac{(|A|^2 - |B|^2)^2}{F^2(t)} \Bigg].
\end{split}
\label{eqn:77}
\end{equation}

The integrated Fisher entropy is
\begin{equation}
I_F(t) = \frac{4m \omega}{\hbar} F(t) \left\langle \left( \frac{H_{\nu}'(\xi)}{H_{\nu}(\xi)} - \xi \right)^2 \right\rangle_{\nu} - \frac{1}{c^4} \frac{\dot{F}(t)^2}{F(t)},
\label{eqn:78}
\end{equation}
where
\begin{equation}
\langle f(\xi)\rangle_{\nu} \equiv \frac{1}{2^{\nu}\nu!\sqrt{\pi}}\int_{-\infty}^{\infty} f(\xi) H_{\nu}^{2}(\xi) e^{-\xi^{2}} d\xi.
\label{eqn:79}
\end{equation}

\begin{figure*}[h!]
\centering
\includegraphics[width=0.9\textwidth]{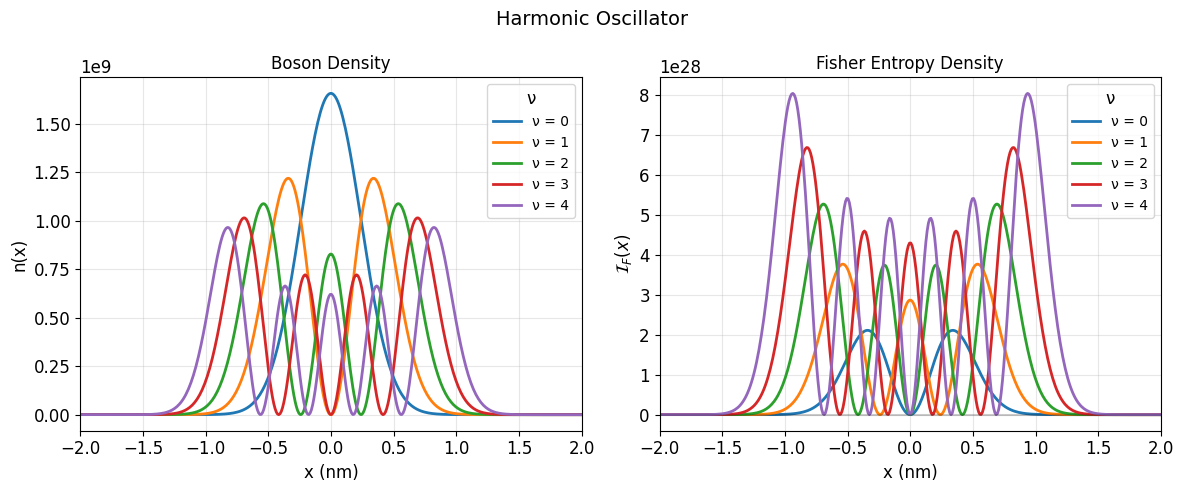}
\caption{\label{fig:1} (Color online) One-dimensional harmonic oscillator. Left: Boson density $n(x)$ for quantum states $\nu = 0, 1, 2, 3, 4$. Right: Corresponding Fisher entropy density $\mathcal{I}_F(x)$. Position $x$ is in nanometers (nm) using the convention from Appendix \ref{Appendix:B}.}
\end{figure*}
\paragraph{Physical Interpretation:}
The Fisher entropy density $\mathcal{I}_{F}$ reveals several key features:
\begin{itemize}
    \item Nodal Peaks: The term $\frac{H_{\nu}'}{H_{\nu}} - \xi$ diverges near the zeros of the Hermite polynomial $H_{\nu}(\xi)$, producing sharp peaks in $\mathcal{I}_{F}$. This indicates that position measurements near wavefunction nodes are highly informative due to the steep gradient in probability density.
    \item Energy Dependence: Higher energy states ($\nu>0$) possess more nodes and spatial oscillations, leading to greater overall Fisher entropy. This reflects increased structural complexity and information content in excited states.
    \item Temporal Interference: For non-stationary superpositions ($A,B\neq0$), the function $F(t)$ modulates $\mathcal{I}_{F}$ in time, demonstrating dynamic redistribution of quantum information due to interference between forward and backward wave components.
\end{itemize}
Thus, $\mathcal{I}_{F}$ maps the "quantum texture" of the state, quantifying where and how wavefunction structure stores measurable information.

This behavior is visually summarized in Fig.~\ref{fig:1}, where the boson density \(n(x)\) and the corresponding Fisher entropy density \(\mathcal{I}_F(x)\) are plotted for the first five quantum states (\(\nu = 0, \ldots, 4\)). The peaks in \(\mathcal{I}_F\) align with the wavefunction nodes, confirming the high information content in regions of steep density gradients.
\subsubsection{Hydrogen Atom}
\label{subsubsec:hydrogen_atom}

For the hydrogen atom, we set $\mathcal{A} = c^{2}$, $A_{0} = \frac{e}{4\pi\epsilon_{0}r}$, $A_{i}=0$, and the mass as $\hat{m}$. After separating variables via
\begin{equation}
\Phi = \Psi(r)Y_{\ell m}(\theta,\phi)\phi(t),
\label{eqn:80}
\end{equation}
with
\begin{equation}
\phi(t) = A e^{i\frac{E}{\hbar}t} + B e^{-i\frac{E}{\hbar}t},
\label{eqn:81}
\end{equation}
the radial function satisfies
\begin{equation}
\frac{1}{r^{2}}\frac{d}{dr}\left(r^{2}\frac{d\Psi}{dr}\right) + g(r)\Psi = 0,
\label{eqn:82}
\end{equation}
where
\begin{equation}
\begin{split}
g(r) = &\frac{E^{2} - \hat{m}^{2}c^{4}}{\hbar^{2}c^{2}} - \frac{\ell(\ell+1)}{r^{2}}\\
&- \frac{2e^{2}E}{4\pi\epsilon_{0}\hbar^{2}c^{2}r} + \frac{e^{4}}{(4\pi\epsilon_{0})^{2}\hbar^{2}c^{2}r^{2}}.
\end{split}
\label{eqn:83}
\end{equation}

The normalized solution is
\begin{equation}
\Psi_{\nu\ell}(r) = \mathcal{D}_{\nu\ell} e^{-\lambda r} r^{\frac{s-1}{2}} L_{\nu}^{s}(2\lambda r),
\label{eqn:84}
\end{equation}
where $L^s_\nu$ are the generalized Laguerre polynomials, with parameters
\begin{align}
s &= \sqrt{(2\ell+1)^{2} - 4\alpha^{2}}, \quad
\lambda = \frac{\hat{m}c\alpha}{\hbar\sqrt{\alpha^{2} + \mathcal{N}^{2}}}, \nonumber \\
\mathcal{N} &= \nu + \frac12 + \frac{s}{2}, \quad
\alpha = \frac{e^{2}}{4\pi\epsilon_{0}\hbar c},
\label{eqn:85}
\end{align}
and quantized energy levels
\begin{equation}
\begin{split}
E_{\nu\ell} = -\frac{\hat{m}c^{2}}{\sqrt{1 + \frac{\alpha^{2}}{\mathcal{N}^{2}}}}, \quad & \nu=0,1,2\dotsc\\
\ell=0,1,\dotsc,\nu-1,
\end{split}
\label{eqn:86}
\end{equation}

The normalization constant is
\begin{equation}
\mathcal{D}_{\nu\ell} = \left[\frac{(2\lambda)^{s+2}\nu!}{\Gamma(\nu + s + 1)(2\nu + s + 1)}\right]^{1/2},
\label{eqn:87}
\end{equation}
and the probability density becomes
\begin{equation}
n = |\mathcal{C}_{\nu\ell m}|^{2} e^{-2\lambda r} r^{s-1} [L_{\nu}^{s}(2\lambda r)]^{2} [P_{\ell}^{m}(\cos\theta)]^{2} G(t),
\label{eqn:88}
\end{equation}
with $G(t) = |\phi(t)|^{2}$ and $\mathcal{C}_{\nu\ell m} = \sqrt{\frac{2\ell+1}{4\pi}\frac{(\ell-m)!}{(\ell+m)!}}\mathcal{D}_{\nu\ell}$.

The phase is
\begin{equation}
\theta = -i\ln\left[\frac{Ae^{i\frac{E}{\hbar}t} + Be^{-i\frac{E}{\hbar}t}}{\sqrt{G(t)}}\right] + m\phi.
\label{eqn:89}
\end{equation}

The geodesic velocity components are
\begin{align}
\pi_{0} &= \frac{E}{\hat{m}c}\frac{|A|^{2} - |B|^{2}}{G(t)} + \frac{1}{4\pi\epsilon_{0}}\frac{e^{2}}{\hat{m}cr}, \nonumber \\
\pi_{r} &= \pi_{\theta} = 0, \quad
\pi_{\phi} = \frac{\hbar m}{\hat{m}}.
\label{eqn:90}
\end{align}

The quantum potential evaluates to [see Eq. (\ref{eqn:12})]
\begin{equation}
U^{Q} = \frac12 \pi_{0}^{2} - \frac{1}{2r^{2}\sin^{2}\theta}\left(\frac{\hbar m}{\hat{m}}\right)^{2} - c^{2}.
\label{eqn:91}
\end{equation}

The divergence of the quantum current is [see Eq. (\ref{eqn:25})]
\begin{equation}
\nabla_{\mu}J^{Q\mu} = \frac{nE}{\hat{m}^{2}c^{3}}\frac{|A|^{2} - |B|^{2}}{G^{2}(t)}\dot{G}(t),
\label{eqn:92}
\end{equation}
while the classical divergence gives
\begin{equation}
\nabla_{\alpha}(J^{\alpha}\pi^{0}) = -\frac{nE}{\hat{m}^{2}c^{3}}\pi^{0}\frac{|A|^{2} - |B|^{2}}{G^{2}(t)}\dot{G}(t),
\label{eqn:93}
\end{equation}
again satisfying equation (\ref{eqn:29}).

The stochastic velocity components are
\begin{align}
u_{0} &= \frac{\hbar}{2\hat{m}c}\frac{\dot{G}(t)}{G(t)}, \nonumber \\
u_{r} &= \frac{\hbar}{2\hat{m}}\left[-2\lambda + \frac{s-1}{r} + 4\lambda\frac{L_{\nu}^{s'}(2\lambda r)}{L_{\nu}^{s}(2\lambda r)}\right], \nonumber \\
u_{\theta} &= -\frac{\hbar}{\hat{m}r^{2}}\frac{P_{\ell}^{m'}(\cos\theta)}{P_{\ell}^{m}(\cos\theta)}\sin\theta, \quad
u_{\phi} = 0.
\label{eqn:94}
\end{align}

The complete velocity $\eta_{\mu} = \pi_{\mu} - iu_{\mu}$ is then
\begin{align}
\eta_{0} &= \frac{E}{\hat{m}c}\frac{|A|^{2} - |B|^{2}}{G(t)} + \frac{1}{4\pi\epsilon_{0}}\frac{e^{2}}{\hat{m}cr} - \frac{i\hbar}{2\hat{m}c}\frac{\dot{G}(t)}{G(t)}, \nonumber \\
\eta_{r} &= -\frac{i\hbar}{2\hat{m}}\left[-2\lambda + \frac{s-1}{r} + 4\lambda\frac{L_{\nu}^{s'}(2\lambda r)}{L_{\nu}^{s}(2\lambda r)}\right], \nonumber \\
\eta_{\theta} &= \frac{i\hbar}{\hat{m}r^{2}}\frac{P_{\ell}^{m'}(\cos\theta)}{P_{\ell}^{m}(\cos\theta)}\sin\theta, \quad
\eta_{\phi} = \frac{\hbar m}{\hat{m}}.
\label{eqn:95}
\end{align}

The chemical potential is
\begin{equation}
\mu_{s} = c^{2}\dot{G}(t),
\label{eqn:96}
\end{equation}
which vanishes in stationary cases ($A=0$ or $B=0$), consistent with thermodynamic equilibrium.

The Fisher entropy density is
\begin{equation}
\mathcal{I}_{F} = \frac{4\hat{m}^{2}}{\hbar^{2}} n\left[-u_{0}^{2} + u_{r}^{2} + u_{\theta}^{2} - \pi_{0}^{2} + \pi_{\phi}^{2}\right],
\label{eqn:97}
\end{equation}
and the non-local part of the integrated Fisher information is
\begin{equation}
\begin{split}
I_{\mathrm{nonlocal}} = & \frac{4\hat{m}^{2}}{\hbar^{2} G(t)}\Bigg[\left(\frac{E}{\hat{m}c}(|A|^{2} - |B|^{2}) + \beta\lambda G(t)\right)^{2}\\
& - \left(\frac{\hbar m}{\hat{m}} G(t)\right)^{2}\Bigg],
\end{split}
\label{eqn:98}
\end{equation}
with $\beta = \frac{e^{2}}{4\pi\epsilon_{0}\hat{m}c}$. The full Fisher information can be written as
\begin{equation}
\begin{split}
I_{F} = & \frac{4\hat{m}^{2}}{\hbar^{2}}\Bigg[\int n\left(-u_{0}^{2} + u_{r}^{2} + u_{\theta}^{2}\right)dV\\ 
&+ \beta^{2}\int n\left(\frac{1}{r} - \left\langle \frac{1}{r}\right\rangle\right)^{2} dV\Bigg].
\end{split}
\label{eqn:99}
\end{equation}
\begin{figure*}[h!]
\centering
\includegraphics[width=0.9\textwidth]{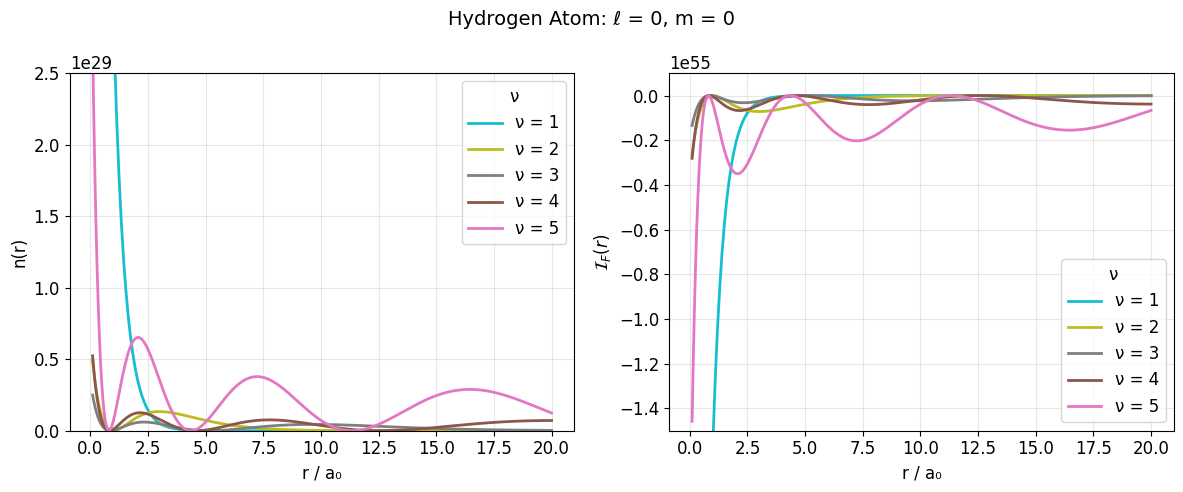}
\caption{\label{fig:2} (Color online) Hydrogen atom with fixed $\ell=0$, $m=0$ (s-states). Left: Density $n(r)$ for principal quantum numbers $\nu = 1, 2, 3, 4, 5$. Right: Fisher entropy $\mathcal{I}_F(r)$. Higher $\nu$ states extend farther from the nucleus and show more radial structure.}
\end{figure*}
\begin{figure*}[h!]
\centering
\includegraphics[width=0.9\textwidth]{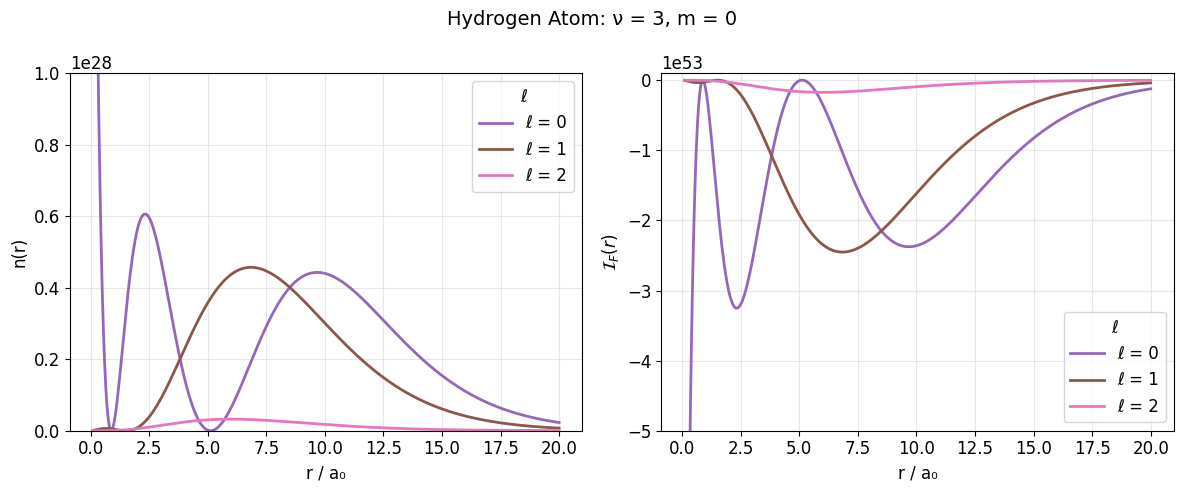}
\caption{\label{fig:3} (Color online) Hydrogen atom with fixed $\nu=3$, $m=0$. Left: Radial density $n(r)$ for angular momentum quantum numbers $\ell = 0, 1, 2$. Right: Fisher entropy $\mathcal{I}_F(r)$. Radial coordinate is in units of Bohr radius $a_0$.}
\end{figure*}
\begin{figure*}[h!]
\centering
\includegraphics[width=0.9\textwidth]{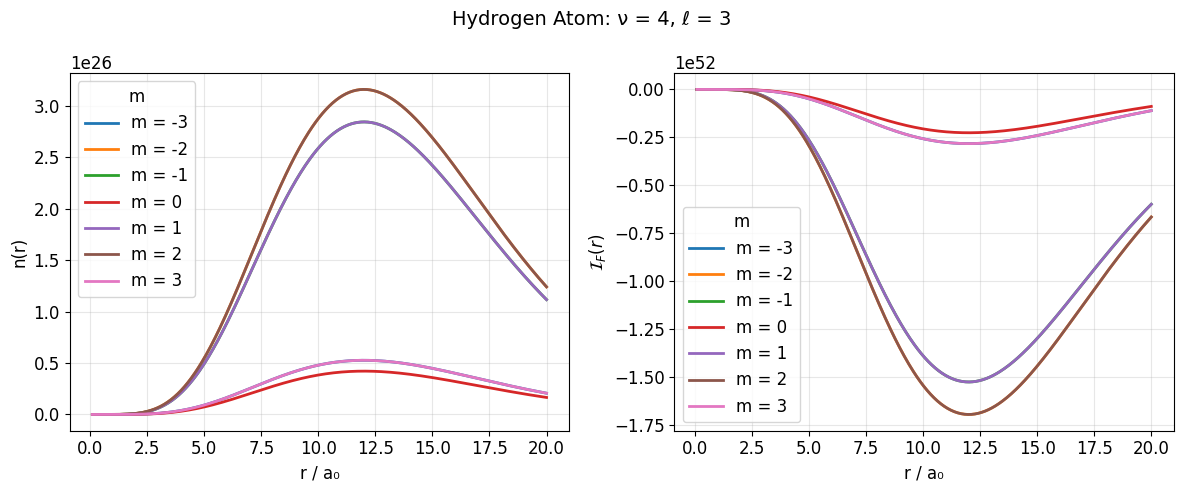}
\caption{\label{fig:4} (Color online) Hydrogen atom with fixed $\nu=4$, $\ell=3$. Left: Density $n(r)$ for magnetic quantum numbers $m = -3, -2, \dots, 3$. Right: Fisher entropy $\mathcal{I}_F(r)$. The angular dependence modulates the radial profiles through spherical harmonics $Y_{\ell m}(\theta,\phi)$ evaluated at $\theta=\pi/4$.}
\end{figure*}
\paragraph{Physical Interpretation:}
The Fisher entropy density \(\mathcal{I}_F\) reveals several key features. First, the term \(\frac{H_{\nu}'(\xi)}{H_{\nu}(\xi)} - \xi\) diverges near the zeros of the Hermite polynomial \(H_{\nu}(\xi)\), producing sharp peaks in \(\mathcal{I}_F\) at the wavefunction nodes. This indicates that position measurements near these nodes are highly informative due to the steep gradients in the probability density. Second, higher energy states (larger \(\nu\)) exhibit more nodes and spatial oscillations, leading to greater overall Fisher entropy. In the relativistic regime, the energy satisfies \(E = \sqrt{\hbar m \omega c^2 (2\nu + 1)}\), so the level spacing grows as \(\sqrt{\nu}\) for large \(\nu\), implying that the information content increases with the square root of the quantum number. This reflects the increased structural complexity of excited states. Third, for non-stationary superpositions (\(A, B \neq 0\)), the function \(F(t) = |\phi(t)|^2\) modulates \(\mathcal{I}_F\) in time, demonstrating a dynamic redistribution of quantum information due to interference between forward and backward wave components. Thus, \(\mathcal{I}_F\) maps the "quantum texture" of the state, quantifying where and how wavefunction structure stores measurable information.
The radial structure of these states and their information content are shown in Figs.~\ref{fig:2}--\ref{fig:4}. For instance, Fig.~\ref{fig:2} illustrates how higher principal quantum numbers \(\nu\) lead to more extended density profiles and richer Fisher entropy structure.
\subsection{Schwarzschild Metric}
\label{subsec:schwarzschild}
The Schwarzschild metric describes the spacetime exterior to a static, spherically symmetric, black hole and serves as a fundamental testing ground for quantum field theory in curved spacetime. Studying the Klein–Gordon field in this geometry allows us to probe how strong gravity modifies quantum dynamics, information content, and thermodynamic behavior. In this section, we apply our formalism to a massive Klein–Gordon field in the Schwarzschild background, with the metric given by \cite{Alcubierre2008}
\begin{equation}
\begin{split}
ds^{2} = & -f(r)c^{2}dt^{2} + f(r)^{-1}dr^{2} + r^{2}d\Omega^2,\\ f(r) = & 1 - \frac{2GM}{c^{2}r}.
\end{split}
\label{eqn:100}
\end{equation}
We consider the massive Klein--Gordon equation, with $\hat{m}$ as the mass
\begin{equation}
(\square - \mu_{\hat{m}}^{2})\Phi = 0,\quad \mu_{\hat{m}} = \frac{\hat{m}c}{\hbar},
\label{eqn:101}
\end{equation}
with $\mathcal{A}=c^{2}$ and $A_{\mu}=0$.

Using the ansatz
\begin{equation}
\begin{split}
\Phi(t,r,\theta,\phi) = & \phi(t)\frac{u(r)}{r} Y_{\ell m}(\theta,\phi),\\
\phi(t) = & A e^{i\omega t} + B e^{-i\omega t},
\end{split}
\label{eqn:102}
\end{equation}
the radial function satisfies
\begin{equation}
f(r)u'' + f'(r)u' + h(r)u = 0,
\label{eqn:103}
\end{equation}
where
\begin{equation}
h(r) = \frac{\omega^{2}/c^{2}}{f(r)} - \frac{\ell(\ell+1)}{r^{2}} - \mu_{\hat{m}}^{2} - \frac{f'(r)}{r}.
\label{eqn:104}
\end{equation}

The physically acceptable solution with exponential decay at infinity is
\begin{equation}
\begin{split}
& u_{\nu\ell} = \\ 
& e^{-\kappa_{\nu}r}\left(1 - \frac{r_{s}}{r}\right)^{\rho_{\nu}} \mathrm{HeunC}\!\left(\alpha_{\nu},\beta_{\nu},0,\delta_{\nu},\eta_{\nu};1 - \frac{r_{s}}{r}\right),\
\end{split}
\label{eqn:105}
\end{equation}
where $\mathrm{HeunC}$ denotes the confluent Heun function \cite{Li2019}. The parameters are
\begin{align}
\alpha_{\nu} &= -2\kappa_{\nu}r_{s}, \nonumber \\
\beta_{\nu} &= 2\rho_{\nu} = 2\sqrt{\frac14 + \ell(\ell+1) - \mu_{\hat{m}}^{2}r_{s}^{2} + 4\kappa_{\nu}^{2}r_{s}^{2}}, \nonumber \\
\delta_{\nu} &= r_{s}^{2}(\tilde{\omega}_{\nu}^{2} + \kappa_{\nu}^{2} - \mu_{m}^{2}), \nonumber \\
\eta_{\nu} &= r_{s}^{2}(\mu_{\hat{m}}^{2} - 2\tilde{\omega}_{\nu}^{2}) - \ell(\ell+1) + \frac12, \nonumber \\
\kappa_{\nu} &= \sqrt{\mu_{\hat{m}}^{2} - \tilde{\omega}_{\nu}^{2}},\quad \tilde{\omega}_{\nu} = \frac{\omega_{\nu}}{c}.
\label{eqn:106}
\end{align}
The quantization condition arises from the requirement that the confluent Heun function reduces to a polynomial, which occurs only for discrete values of the parameter $\omega_\nu$. This polynomial termination condition yields the discrete energy spectrum
\begin{equation}
\omega_{\nu}^{2} = \mu_{\hat{m}}^{2}c^{2} - \kappa_{\nu}^{2} = \left(\frac{\hat{m}c^{2}}{\hbar}\right)^{2} - \kappa_{\nu}^{2}c^{2},
\label{eqn:107}
\end{equation}
corresponding to bound states of the bosonic field in the Schwarzschild background \cite{Li2019}. The integer $\nu = 0,1,2,\dots$ (see Eq. (\ref{eqn:121})) labels the radial excitation levels.\\

The full wavefunction is
\begin{equation}
\Phi_{\nu\ell m}(x_{\mu}) = \mathcal{N}_{\nu\ell}\left(Ae^{i\omega_{\nu}t} + Be^{-i\omega_{\nu}t}\right)\frac{u_{\nu\ell}(r)}{r} Y_{\ell m}(\theta,\phi),
\label{eqn:108}
\end{equation}
with normalization constant ensuring
\begin{equation}
\int \sqrt{-g}|\Phi_{\nu\ell m}|^{2} d^{3}x = 1,\quad \mathcal{N}_{\nu\ell}^{-2} = \int_{r_{s}}^{\infty}|u_{\nu\ell}(r)|^{2} dr.
\label{eqn:109}
\end{equation}

The phase is
\begin{equation}
\theta = -i\ln\left[\frac{Ae^{i\omega t} + Be^{-i\omega t}}{\sqrt{H(t)}}\right] + m\phi,\quad H(t) = |\phi(t)|.
\label{eqn:110}
\end{equation}

The geodesic velocity components are
\begin{align}
\pi_{0} &= \frac{\hbar\omega_{\nu}}{\hat{m}c}\frac{|A|^{2} - |B|^{2}}{H(t)}, \nonumber \\
\pi_{r} &= \pi_{\theta} = 0, \quad
\pi_{\phi} = \frac{\hbar m}{\hat{m}}.
\label{eqn:111}
\end{align}

The divergences of the quantum and classical currents are
\begin{align}
\nabla_{\mu}J^{Q\mu} &= \frac{n}{f(r)^{2}c^{5}}\left(\frac{\hbar\omega_{\nu}}{\hat{m}c}\right)^{2}(|A|^{2} - |B|^{2})^{2}\frac{H'(t)}{H(t)^{3}}, \label{eqn:112} \\
\nabla_{\mu}(J^{\mu}\pi^{0}) &= -\frac{n}{f(r)^{2}c^{5}}\left(\frac{\hbar\omega_{\nu}}{\hat{m}c}\right)^{2}(|A|^{2} - |B|^{2})^{2}\frac{H'(t)}{H(t)^{3}}, \label{eqn:113}
\end{align}
again satisfying the energy balance equation (\ref{eqn:29}).

The stochastic velocity components are
\begin{align}
u_{0} &= \frac{\hbar}{2\hat{m}c}\frac{\dot{H}(t)}{H(t)}, \nonumber \\
u_{r} &= \frac{\hbar}{2\hat{m}}\left[\frac{u_{\nu\ell}'(r)}{u_{\nu\ell}(r)} - \frac{1}{r}\right], \nonumber \\
u_{\theta} &= -\frac{\hbar}{\hat{m}r^{2}}\frac{Y_{\ell m}'(\theta,\phi)}{Y_{\ell m}(\theta,\phi)}\sin\theta, \quad
u_{\phi} = 0.
\label{eqn:114}
\end{align}

The complete velocity is
\begin{align}
\eta_{0} &= \frac{\hbar\omega_{\nu}}{\hat{m}c}\frac{|A|^{2} - |B|^{2}}{H(t)} - \frac{i\hbar}{2\hat{m}c}\frac{\dot{H}(t)}{H(t)}, \nonumber \\
\eta_{r} &= -\frac{i\hbar}{2\hat{m}}\left[\frac{u_{\nu\ell}'(r)}{u_{\nu\ell}(r)} - \frac{1}{r}\right], \nonumber \\
\eta_{\theta} &= \frac{i\hbar}{\hat{m}r^{2}}\frac{Y_{\ell m}'(\theta,\phi)}{Y_{\ell m}(\theta,\phi)}\sin\theta, \quad
\eta_{\phi} = \frac{\hbar m}{\hat{m}}.
\label{eqn:115}
\end{align}

The chemical potential evaluates to
\begin{equation}
\mu_{s} = c^{2}H'(t)I_{f},\quad I_{f} = \int_{r_{s}}^{\infty}\frac{r}{f(r)} u(r) dr,
\label{eqn:116}
\end{equation}
representing energy transfer due to temporal variations in boson density, enhanced near the horizon where $f(r)\to0$.

The thermodynamic coupling term vanishes:
\begin{equation}
\mathcal{T} = 0,
\label{eqn:117}
\end{equation}
since $\partial_{t}(\sqrt{\gamma}/N) = 0$ for the static Schwarzschild metric, reflecting no energy exchange with the stationary spacetime geometry.

The Fisher entropy density is
\begin{equation}
\begin{split}
\mathcal{I}_{F} = &  \frac{4\hat{m}^{2}}{\hbar^{2}} n\Bigg[-\frac{u_{0}^{2}}{f(r)} + f(r)u_{r}^{2} + \frac{u_{\theta}^{2}}{r^{2}}\\
&- \frac{\pi_{0}^{2}}{f(r)} + \frac{\pi_{\phi}^{2}}{r^{2}\sin^{2}\theta}\Bigg],
\end{split}
\label{eqn:118}
\end{equation}
and the integrated Fisher information becomes
\begin{equation}
\begin{split}
& I_{F} =  \frac{4\hat{m}^{2}}{\hbar^{2}}\int \sqrt{-g} n\Bigg[-\frac{1}{f(r)}\left(\frac{\hbar}{\hat{m}}\frac{\dot{H}}{H}\right)^{2} \\
&+ f\left[\frac{\hbar}{2\hat{m}}\left(\frac{u'}{u} - \frac{1}{r}\right)\right]^{2} + \frac{1}{r^{2}}\left(\frac{\hbar}{\hat{m}r^{2}}\frac{Y'}{Y}\sin\theta\right)^{2}\Bigg] d^{3}x.
\end{split}
\label{eqn:119}
\end{equation}
\begin{figure*}[h!]
\centering
\includegraphics[width=0.9\textwidth]{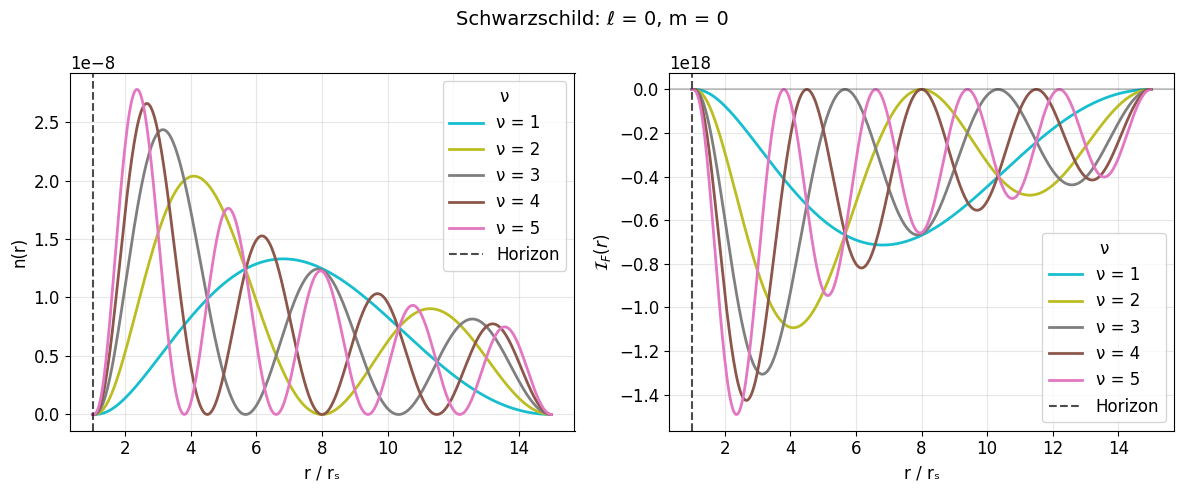}
\caption{\label{fig:5} (Color online) Klein-Gordon field in Schwarzschild geometry with fixed $\ell=0$, $m=0$ (spherically symmetric modes). Left: Density $n(r)$ for radial quantum numbers $\nu = 1, 2, 3, 4, 5$. Right: Fisher entropy $\mathcal{I}_F(r)$. Higher $\nu$ states decay faster and exhibit more spatial oscillations.}
\end{figure*}
\begin{figure*}[h!]
\centering
\includegraphics[width=0.9\textwidth]{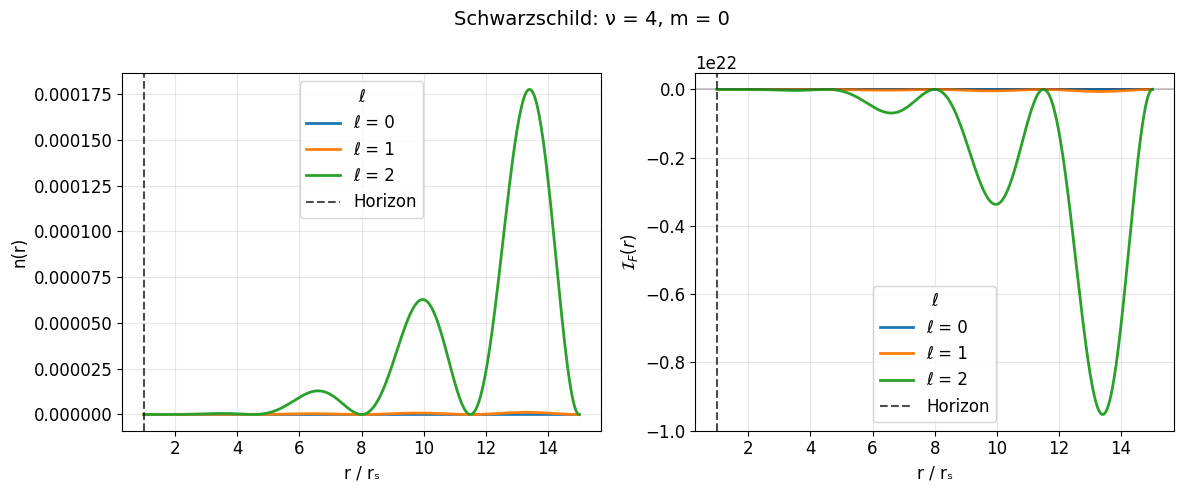}
\caption{\label{fig:6} (Color online) Klein-Gordon field in Schwarzschild geometry with fixed $\nu=4$, $m=0$. Left: Density $n(r)$ for angular momentum numbers $\ell = 0, 1, 2, 3$. Right: Fisher entropy $\mathcal{I}_F(r)$. Radial coordinate is in units of Schwarzschild radius $r_s = 2GM/c^2$. The dotted vertical line marks the event horizon. Original numerical solutions.}
\end{figure*}
\paragraph{Physical Interpretation:}

The Fisher entropy density reveals several key features in the Schwarzschild metric. First, the factor \(1/f(r)\) diverges as \(r \to r_s\), dramatically amplifying \(\mathcal{I}_F\) near the event horizon. This indicates that quantum field structure becomes extremely sharp and informative in strong gravitational regions, demonstrating how gravity enhances quantum fluctuations \cite{Hawking1975,Escobar-Aguilar2025}. Second, terms involving \(u'/u\) and \(u''/u\), governed by the Heun function solution, reflect how spacetime geometry shapes quantum information differently from flat-space cases \cite{Li2019}. Third, the entropy density reflects the discrete "black hole atom" energy levels, localized where the wavefunction has peaks and nodes, now distorted by strong gravity. Finally, for \(\ell \geq 2\), centrifugal barriers move the equilibrium distance where \(\mathcal{I}_F\) vanishes farther from the horizon, showing how angular momentum modifies information distribution in curved spacetime.
This behavior reveals that the most complex and information-rich region is not at the center of the bound state, but is dramatically compressed and amplified near the event horizon itself. The concentration of quantum information near the horizon is consistent with the holographic principle~\cite{tHooft2009, Bekenstein1973}, which suggests that information in gravitational systems is encoded on boundary surfaces~\cite{Hawking1975}, highlighting the profound interplay between gravity and quantum information.
In the Schwarzschild metric, the Fisher entropy density reveals how intense gravity sculpts quantum fields, compressing and amplifying quantum information near the event horizon. This provides an information-theoretic perspective on black hole--quantum field interactions, complementing traditional thermodynamic approaches \cite{Hawking1975, tHooft2009}.
The gravitational distortion of quantum fields is quantified in Figs.~\ref{fig:5} and \ref{fig:6}. Fig.~\ref{fig:5} shows the boson density \(n(r)\) and Fisher entropy \(\mathcal{I}_F(r)\) for spherically symmetric modes (\(\ell = 0\)) with increasing radial quantum number \(\nu\). The amplification of \(\mathcal{I}_F\) near the event horizon (\(r \to r_s\)) is clearly visible.
\section{Conclusion}
\label{sec:conclusion}

In this work, we have developed a refined thermodynamic description for a zero-temperature boson gas in curved spacetime. Building upon the hydrodynamic formulation of the Klein-Gordon-Maxwell system within the ADM \(3+1\) formalism \cite{Alcubierre2008}, we have derived a dual formulation that separates energy transport from information conservation, providing deeper physical insight.

\begin{enumerate}
    \item \textbf{The Energy Balance Equation (\ref{eqn:29})}: This equation, $\nabla_{\mu}\mathcal{J}^{\mu} + n\nabla^{0}\mathcal{A} = 0$, governs the flux of total energy (kinetic, quantum, and electromagnetic) through spacetime and its coupling to temporal changes in the interaction potential. It represents the thermodynamic first law from the perspective of the spacetime foliation, extending previous work on bosonic systems in curved spacetime \cite{Matos2019, Chavanis2017_b}.
    
    \item \textbf{The Information-Theoretic Constraint (\ref{eqn:60})}: This equation, $\mathcal{I}_F = 2\square n - \frac{8m^2}{\hbar^2} n\mathcal{A}$, establishes a fundamental link between dynamics and information. It shows that the Fisher information entropy $\mathcal{I}_F$---a measure of the ``structural information'' or ``quantumness'' embedded in the wavefunction---is determined by the balance between the spacetime d'Alambertian of the density and the system's self-interactions. This acts as a law of information conservation in curved spacetime, connecting quantum dynamics with information theory \cite{Frieden2004, DnesPetz2008}. 
\end{enumerate}

This dual framework elegantly separates the dynamics of energy transport from the constraints of information conservation. Furthermore, the introduction of the stochastic velocity $u_\mu$ and the associated Fisher entropy provides a consistent mathematical framework that separates the diffusive and conservative components of the dynamics, leading to the information-theoretic constraint (\ref{eqn:60}).

The formalism was rigorously tested in specific cases---the quantum harmonic oscillator \cite{Schiff1968}, the hydrogen atom \cite{Schiff1968}, and the Klein-Gordon field in Schwarzschild geometry \cite{Li2019, Hawking1975}---confirming the consistency of the approach and illustrating the local, reversible energy exchange between classical and quantum currents.

By unifying concepts from general relativity \cite{Alcubierre2008}, thermodynamics \cite{Matos2019}, and information theory \cite{Frieden2004}, this work provides a powerful foundation for modeling relativistic bosonic systems, such as boson stars \cite{Chavanis2011} and scalar field dark matter \cite{Matos2001, Guzmn2000}, offering new insights into the interplay of energy, information, and gravity.
\section*{Acknowledgments}
Jorge Meza-Domínguez thanks SECIHTI-M\'exico for the doctoral fellowship No. 1235731.\\
This work was also partially supported by SECIHTI M\'exico under grants SECIHTI CBF-2025-G-1720 and CBF-2025-G-176. The authors are gratefully for the computing time granted by LANCAD and CONACYT in the Supercomputer Hybrid Cluster "Xiuhcoatl" at GENERAL COORDINATION OF INFORMATION AND COMMUNICATIONS TECHNOLOGIES (CGSTIC) of CINVESTAV. URL: http://clusterhibrido.cinvestav.mx/ and to Hector Oliver Hernandez for his help with the code installations.\\
All figures in this paper are original creations produced specifically for this study. The numerical implementation was developed from first principles based on our theoretical framework, ensuring consistency between analytical derivations and visual representations.
\appendix

\section{The Confluent Heun Function (HeunC) and Identities}

\subsection{Definition of the Confluent Heun Function}

The Confluent Heun function, denoted as $\mathrm{HeunC}(\alpha, \beta, \gamma, \delta, \eta; z)$, is a solution to the confluent Heun differential equation~\cite{Li2019}
\begin{equation}
\begin{split}
\frac{d^2w}{dz^2} &+ \left( \frac{\beta+1}{z} + \frac{\gamma+1}{z-1} + \alpha \right) \frac{dw}{dz}\\
&+ \left( \frac{\mu}{z} + \frac{\nu}{z-1} \right) w = 0,
\end{split}
\end{equation}
where the parameters are related to the standard notation by
\begin{align*}
\mu &= \frac{1}{2}(\alpha - \beta - \gamma + \alpha\beta - \beta\gamma) - \eta, \\
\nu &= \frac{1}{2}(\alpha + \beta + \gamma + \alpha\gamma + \beta\gamma) + \delta + \eta.
\end{align*}

\subsection{Polynomial Solutions and Quantization Condition}

For specific parameter values, the Confluent Heun function reduces to a polynomial. The polynomial termination condition occurs when:

\begin{equation}
\frac{\delta}{\alpha} + \frac{\beta + \gamma + 2}{2} + n = 0, \quad n = 0, 1, 2, \ldots,
\label{eqn:121}
\end{equation}
this quantization condition yields discrete energy spectra in physical applications, such as the bound states in the Schwarzschild metric analyzed in Section 6.2, providing the mathematical foundation for quantum states in curved spacetime~\cite{Li2019}.

\subsection{Asymptotic Behavior and Regularity}

Near the regular singular point $z = 0$, the HeunC function behaves as
\begin{equation}
\begin{split}
\mathrm{HeunC}(\alpha, \beta, \gamma, \delta, \eta; z)\\
\sim 1 - &\frac{(\alpha\beta - \beta\gamma - 2\eta + \alpha)}{2(\beta+1)}z + O(z^2).
\end{split}
\end{equation}
Near \(z = 1\) (which corresponds to \(r \to \infty\) in the physical coordinate), the behavior is given by
\begin{equation}
\mathrm{HeunC}(\alpha, \beta, \gamma, \delta, \eta; z) \sim C_1(1-z)^{-\gamma} + C_2,
\end{equation}
This asymptotic analysis is crucial for understanding the behavior of quantum fields near the event horizon in black hole spacetimes~\cite{Hawking1975}. The coordinate transformation \(z = 1 - r_s/r\) maps the horizon \(r = r_s\) to \(z = 0\) and spatial infinity \(r \to \infty\) to \(z = 1\). Thus, the behavior at \(z = 1\) (infinity) determines the regularity conditions that, together with the behavior at \(z = 0\) (the horizon), yield the discrete bound state spectrum.
\subsection{Special Cases and Relationships}

The Confluent Heun function generalizes many special functions~\cite{Li2019}:
\begin{itemize}
\item When $\alpha = 0$, it reduces to the hypergeometric function ${}_2F_1$, connecting to standard quantum mechanical problems.
\item For specific parameter choices, it becomes the Mathieu, Bessel, or Laguerre functions, demonstrating its versatility in physical applications.
\item In the Schwarzschild metric application (Section 6.2), it appears in the form.
\begin{equation}
\mathrm{HeunC}\left(\alpha_\nu, \beta_\nu, 0, \delta_\nu, \eta_\nu; 1-\frac{r_s}{r}\right),
\end{equation}
with parameters defined in equations (\ref{eqn:105})-(\ref{eqn:107}) of the main text, providing exact solutions for bosonic fields in strong gravitational fields.
\end{itemize}

\subsection{Wronskian and Derivative Identities}

The Wronskian of two independent solutions is given by
\begin{equation}
\mathcal{W}[\mathrm{HeunC}_1, \mathrm{HeunC}_2] = z^{-\beta-1}(z-1)^{-\gamma-1}e^{-\alpha z},
\end{equation}
the derivative can be expressed as
\begin{equation}
\begin{split}
&\frac{d}{dz}\mathrm{HeunC}(\alpha, \beta, \gamma, \delta, \eta; z) =\\
&\frac{\beta}{z}\mathrm{HeunC}(\alpha, \beta, \gamma, \delta, \eta; z)\\
&-\frac{\mu}{z}\mathrm{HeunC}(\alpha, \beta+1, \gamma, \delta, \eta'; z),
\end{split}
\end{equation}
where $\eta'$ is a modified parameter. These identities are essential for constructing complete sets of solutions in quantum mechanical problems.

\subsection{Normalization and Orthogonality}

For polynomial solutions, the normalization integral:
\begin{equation}
\int \mathrm{HeunC}_m(z)\mathrm{HeunC}_n(z)w(z)dz = \delta_{mn},
\end{equation}
involves a weight function $w(z)$ that depends on the specific parameter values and ensures orthogonality of the eigenfunctions in quantum mechanical applications, providing the mathematical basis for the probabilistic interpretation of quantum states in curved spacetime~\cite{Schiff1968}.
\section{Numerical implementation}
\label{Appendix:B}

All numerical computations and visualizations were performed using original Python code. The plots presented in Figs.~\ref{fig:1}--\ref{fig:6} are based on direct evaluation of the analytical expressions derived in this work, specifically:
\begin{itemize}
    \item Harmonic oscillator: Eqs.~(\ref{eqn:66})-(\ref{eqn:67}) for wavefunctions, Eq.~(\ref{eqn:60}) for Fisher entropy.
    \item Hydrogen atom: Eqs.~(\ref{eqn:84})-(\ref{eqn:86}) for bound states, with spherical harmonics $Y_{\ell m}$.
    \item Schwarzschild field: Numerical solutions based on Eqs. (\ref{eqn:105})-(\ref{eqn:107}) (evaluating the confluent Heun function numerically), with gravitational redshift included.
\end{itemize}
Physical parameters: electron mass $m_e = 9.11\times10^{-31}$~kg, Bohr radius $a_0 = 5.29\times10^{-11}$~m, solar mass $M_\odot = 1.99\times10^{30}$~kg. The radial coordinate in black hole plots is normalized to the Schwarzschild radius $r_s = 2GM/c^2$.

\bibliographystyle{ieeetr}      
\bibliography{GEB}
\end{document}